\documentclass{article}

\usepackage{graphicx}

\def\ligne#1 {\hbox to\hsize{#1}}
\def\leurre{\noindent\leftskip 0pt\footnotesize\baselineskip 10pt\parindent 0pt}

\newtheorem{fig}{{\sc Figure}}
\newtheorem{defn}{{\sc Definition}}
\newtheorem{tab}{{\sc Table}} 

\newtheorem{thm}{{\sc Theorem}}

\begin{document}
% \title[short header title]{title}
\ligne{\hfill
\bf\Large A weakly universal cellular automaton in the\hfill} 
\ligne{\hfill \bf\Large pentagrid with five states\hfill}

\ligne{\hfill Maurice {\sc Margenstern}\hfill}
%\author[2]{first name 2}{last name 2}
\vskip 30pt
% addresses are automatically numbered
\ligne{\hfill Universit\'e de Lorraine\hfill}
\ligne{\hfill LITA, EA 3097,\hfill} 
\ligne{\hfill Campus du Saulcy,\hfill}
\ligne{\hfill 57045 Metz, C\'edex 1, France\hfill}
\ligne{\hfill {\it e-mail}: {\tt maurice.margenstern@univ-lorraine.fr, margenstern@gmail.com}
\hfill}
%\address{...\\...\\
%         ..., country}
%        {e-mail 1}
\vskip 20pt
{\bf Abstract} {\it\small
In this paper, we construct a cellular automaton on the pentagrid which is planar,
weakly universal and which have five states only. This result much improves the best result
which was with nine states.}

{\bf Keywords} {\it cellular automata, universality, tilings, hyperbolic geometry.}

\vskip 10pt

%\maketitle

%\newpage

\section{Introduction}

   In this paper, we construct a weakly universal cellular automaton on the pentagrid,
see Theorem~\ref{univ5} at the end of the paper. Two papers, \cite{fhmmTCS,mmsyPPL}
already constructed such a cellular automaton, the first one with 22~states, the second one with
9~states. In this paper, the cellular automaton we construct has five states only.
It uses the same principle of simulating a register machine through a railway circuit, but
the implementation takes advantage of new ingredients introduced by the author in his quest
to lower down the number of states, see~\cite{mmbook3}. 
The reader is referred to~\cite{mmbook1,mmbook2,mmbook3} for an introduction to hyperbolic
geometry turned to the implementation of cellular automata in this context. A short introduction
can also be found in~\cite{mmDMTCS}. However, it is not required to be an expert in hyperbolic 
geometry in order to read this paper.

Section~\ref{univ} reminds the definition we take for 
weak universality. Section~\ref{implement} is devoted to the proof of Theorem~\ref{univ5}.
Section~\ref{implement} of the paper. In that section, Subsection~\ref{railway} reminds
the basic model used in the paper, Subsection~\ref{inpenta} explains its implementation in
the pentagrid, the tiling $\{5,4\}$ of the hyperbolic plane, Subsection~\ref{scenar} explains
the scenario of the simulation performed by the automaton proving Theorem~\ref{univ5} and 
Subsection~\ref{rules} gives the rules of the cellular automaton.

\section{Universality and weak universality}
\label{univ}

    Universality is a well know notion in computer science. However, the single word 
'universality' is understood in different ways, sometimes somehow divergent.

    Let us go back to the definition.

\vskip 5pt
\vtop{
\begin{defn}\label{univdef}
Let~$\cal K$ be a class of processes. Say that $\cal K$ 
{\bf possesses a universal element}~$U$ if, a finite alphabet~$A$ being fixed once and for all,
there is an encoding~$c$ of $\cal K$ elements into the words on~$A$ and of the data for
a $\cal K$-element into the words on~$A$ such that for all element~$\chi$ of~$\cal K$ and for
all data~$d$ of~$\chi$, $U$~applied to $(c(\chi),c(d))$ ends its computation if and only if
$\chi$ ends its own one when it is applied to~$d$ and, in that case, if
\hbox{$U(c(\chi),c(d))=c(\chi(d))$}.
\end{defn}
}

\noindent
We also say that $U$ {\bf simulates}~$\chi$ or that~$\chi$ is {\bf smulated} by~$U$.
If $\cal K$ possesses a universal element, we also say that $\cal K$ possesses the 
property of being universal. We get again the standard definition of universal Turing machine,

   In the above definition, therei are four elements. Data~$\chi$ and~$d$, the encoding of~$c$ 
and the universal element~$U$. From the definition itself, the notion of encoding is an essential 
feature. Indeed, among the elements of~$\cal K$, it is not difficult to construct some of them whose
encoding is bigger than that of~$U$. Indeed, it may be assumed that the encoding is an increasing 
function in this sense that if $\chi$ and~$\xi$ are elements of~$\cal K$ transforming words 
on~$A$ onto words on~$A$, then \hbox{$c(\xi_\circ\,\!\chi)> c(\chi),c(\xi)$}. 
Consequently, encoding the elements of~$\cal K$ into a fixed alphabet is an essential feature.
It allows~$U$ to simulate objects which are bigger than itself. Of course, changing the encoding
may result in a change on the computation of~$U$ which may then be either faster or slower.
At last, when~$U$ stops its computation, the result is an encoding of the result of the
element of~$\cal K$ simulated by~$U$.

   Now, since a few decades, these three items: the data, the encoding and the result are not
always considered in the same way. From the definition, $d$~is finite, as a word on~$A$;
when the computation of~$\chi$ on~$d$ stops, that of~$U$ on~$c(\chi)$ and~$c(d)$ also stops.
Now, whether this latter condition is observed or not, it happens that when~$U$ is applied 
to~$c(\chi)$ and~$c(d)$, it does not yield~$c(d)$ when~$\chi$ completes its computation on~$d$,
but something else, call it~$e(d)$, where~$e$ can be considered as another encoding of~$d$,
$e$~being also fixed once and for all.
Indeed, \hbox{$U(c(\chi),c(d))=c(\chi(d))$} can also be rewritten 
\hbox{$c^{-1}(U(c(\chi),c(d)))=\chi(d)$}, so that $\chi(d)$ is restored by
{\bf decoding} $U(c(\chi),c(d)))$. Introducing~$e$ consists in accepting that the decoding
funcion can be independent from the encoding one.

   Now, the conditions of finiteness on~$d$ and on the computation of~$U$ when that of~$chi$
on the considered data stops are not always observed. When all conditions of 
Definition~\ref{univdef} are observed we say that $U$~is {\bf strongly universal}.
In this definition of strong universality, it is not required that the decoding be the inverse
function of the encoding. However, it is required that~$e$ and~$c$ belong to comparable classes
of complexity. Specifically, it is required that there is a primitive recursive function~$u$
such that for any~$d$, \hbox{$c(d),e(d)\leq u(\vert d\vert)$}, where $\vert d\vert$ 
is the size of~$c(d)$, {\it i.e.} the number of symbols in~$c(d)$.

   When the conditions of strong universality are not observed, we say that~$U$ is 
{\bf weakly universal}. In case $d$~is in some sense infinite, it is required that
$c(d)$, which is an infinite word be of the form \hbox{$u^{\ast}wv^{\ast}$}, where
$u$, $v$ and~$w$ are words on~$A$. The repeated words~$u$ and~$v$ are called the periodic
patterns and it is not required that \hbox{$u=v$}. Note that during the computation,
we may consider that each step~$t$ of~$U$ works on something of the form
\hbox{$u^{\ast}w_tv^{\ast}$}. It is this situation that we shall consider, with this
difference that we work in a $2D$-space and so, accordingly, we require a condition of
periodicity restricted to the outside of a big enough disc, this condition being able to
involve two different periodic patterns.

\section{The scenario of our simulation}
\label{implement}

   Most of the cellular automata in hyperbolic spaces I constructed, myself or with
a co-author, apply the same model of computation. We implement a railway circuit devised
by Ian Stewart, see~\cite{stewart} which we rework in order to simulate a register machine.
This general scenario is described in detail in~\cite{mmbook2,mmbook3}. Here,
we simply give the guidelines in order to introduce the changes which are specific to this
implementation.

\subsection{Basic features}
\label{railway}

   The railway circuits consists of tracks, crossings and switches, and a single locomotive
runs over the circuit. The tracks are pieces
of straight line or arcs of a circle. In the hyperbolic context, we shall replace these
features by assuming that the tracks travel either on {\bf verticals} or {\bf horizontals}
and we shall make it clear a bit later what we call by these words. The crossing is
an intersection of two tracks, and the locomotive which arrives at an intersection 
by following a track goes on by the track which naturally continues the track through which
it arrived. Again, later we shall make it clear what this natural continuation is.
Below, Figure~\ref{switches} illustrates the switches and Figure~\ref{element} illustrates
the use of the switches in order to implement a memory element which exactly contains 
one bit of information.

   The three kinds of switches are the {\bf fixed switch}, the {\bf flip-flop}
and the {\bf memory switch}. In order to understand how the switches work, notice that in
all cases, three tracks abut the same point, the {\bf centre} of the switch. On one side
switch, there is one track, say~$a$, and on the other side, there are two tracks, call 
them~$b$ and~$c$. When the locomotive arrive through~$a$, we say that it is an {\bf active}
crossing of the switch. When it arrives either through~$b$ or~$c$, we say that it is a
{\bf passive} crossing. 

   In the fixed switch, in an active passage, the locomotive is sent either always to~$b$
or always to~$c$, we say that the {\bf selected track} is always~$b$ or it is always~$c$. 
In the passive crossing, the switch does nothing, the locomotive leaves
the switch through~$a$. In the flip-flop, passive crossings are prohibited: the circuit
must be managed in such a way that a passive crossing never occurs at any flip-flop.
During an active passage, the selected track is changed just after the passage of the locomotive:
if it was~$b$, $c$ before the crossing, it becomes~$c$, $b$ respectively after it.
In the memory switch, both active and passive crossing are allowed. The selected tracks also
may change and the change is dictated by the following rule: after the first crossing, only
in case it is active, the selected track is always defined as the track taken by the locomotive
during its last passive crossing of the switch. The selected track at a given switch defines
its {\bf position}.

\vtop{
\vspace{5pt}
\ligne{\hfill
\includegraphics[scale=1.0]{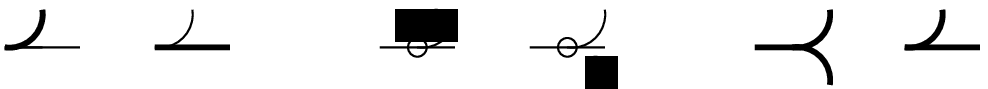}
\hfill}
\vspace{-5pt}
\begin{fig}\label{switches}
\leurre
Les aiguillages du circuit ferroviaire. De gauche \`a droiite~: le fixe, la bascule
et le m\'emorisant.
\end{fig}
\vspace{10pt}
}

   The current configuration of the circuit is the position of all the switches of the circuit.
Note that it may be coded in a finite word, even if the circuit is infinite, as at each time,
only finitely many switches have been visited by the locomotive.

   Figure~\ref{element} illustrates how a flip-flop and a memory switch can be coupled 
in order to make a one bit memory element.

\vtop{
\vspace{-5pt}
\ligne{\hfill
\includegraphics[scale=0.81]{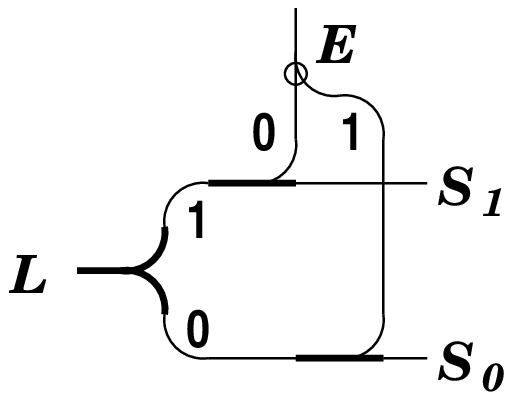}
\hfill}
\ligne{\hfill
\includegraphics[scale=0.54]{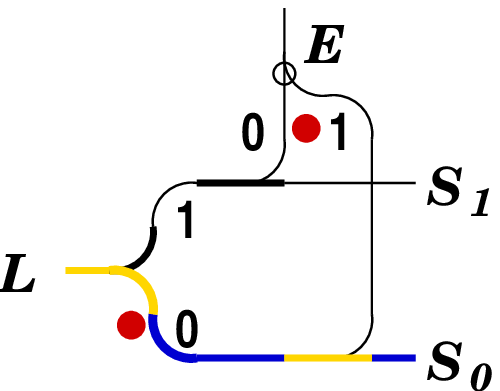}
\includegraphics[scale=0.54]{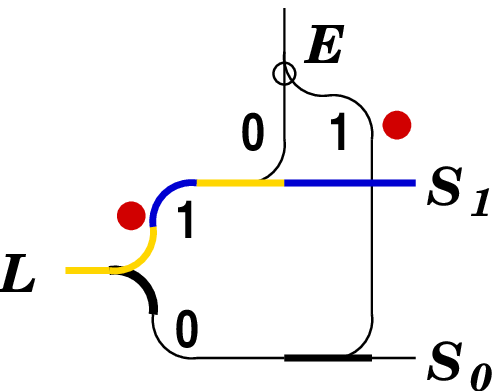}
%\hfill}
%\vspace{4pt}
%\ligne{
\hfill
\includegraphics[scale=0.54]{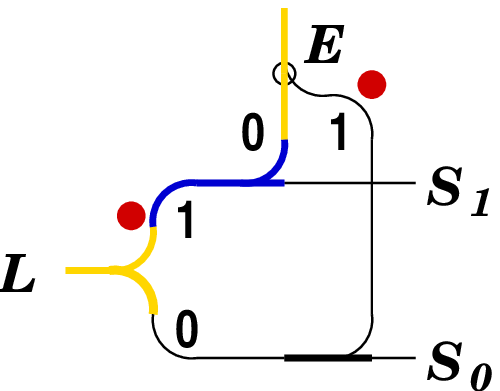}
\includegraphics[scale=0.54]{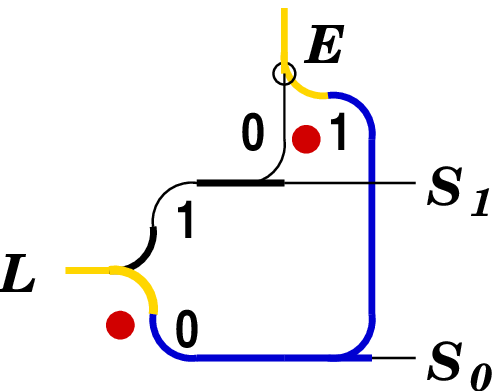}
\hfill}
\vspace{-5pt}
\begin{fig}\label{element}
\leurre
L'\'el\'ement de base du circuit. Deuxi\`eme ligne~: 
\vskip 0pt
\vspace{-2pt}
Deux premiers dessins~: 
lecture de l'\'el\'ement. Deux derniers dessins~: \'ecriture de l'\'el\'element.
\end{fig}
%\vspace{10pt}
}

\subsection{In the pentagrid}
\label{inpenta}

   As mentionned in the introduction, the first weakly universal cellular automaton on the
pentagrid was done by the author and a co-author, see~\cite{fhmmTCS}. The paper implements
the solution sketchily mentioned in Subsection~\ref{railway} with 22 states. In the next paper 
about a weakly universal cellular automaton on the pentagrid, see~\cite{mmsyPPL}, the same model
is implemented with 9~states. The difference with the former paper is that in the second paper,
the cell which is at the centre of the switch has the same colour as another cell of the track.
The centre of the switch is signalized by the neighbouring of the centre.

\subsubsection{Former implementations}

   Figure~\ref{hca54stables} illustrates the implementation of the crossing and of the
switches performed in~\cite{mmsyPPL}, showing in particular, the feature at which we just pointed.
Figure~\ref{hca54voies} shows the implementation of the verticals, second row in the figure,
and of the horizontal, first row. Both these figures show how to implement the basic element
of Figure~\ref{element} in Subsection~\ref{railway}. Figure~\ref{hca54element} show a global view
of how the tree structure of the tiling can be used to implement a basic element in the
pentagrid.

\def\WW{\hbox{\tt W}{}}
\def\BB{\hbox{\tt B}{}}
\def\GG{\hbox{\tt G}{}}
\def\RR{\hbox{\tt R}{}}
\def\DB{\hbox{\tt B$_D$}{}}
\def\DG{\hbox{\tt G$_D$}{}}
\def\LG{\hbox{\tt G$_L$}{}}
\def\DR{\hbox{\tt R$_D$}{}}
\def\XX{\hbox{\tt X}{}}

\ifnum 1=0 {
\vtop{
\vspace{-5pt}
\ligne{\hfill
\includegraphics[scale=0.3]{hca_univ_global.ps}
\hfill}
\vspace{-30pt}
\begin{fig}\label{hca_univglobal}
\leurre
Aper\c cu global de l'implantation de l'exemple-jouet de la figure~{\rm\ref{sequencer}}
dans la pentagrille.
\end{fig}
\vspace{10pt}
}
} \fi

\vtop{
\vspace{-5pt}
\ligne{\hskip 5pt
\includegraphics[scale=0.4]{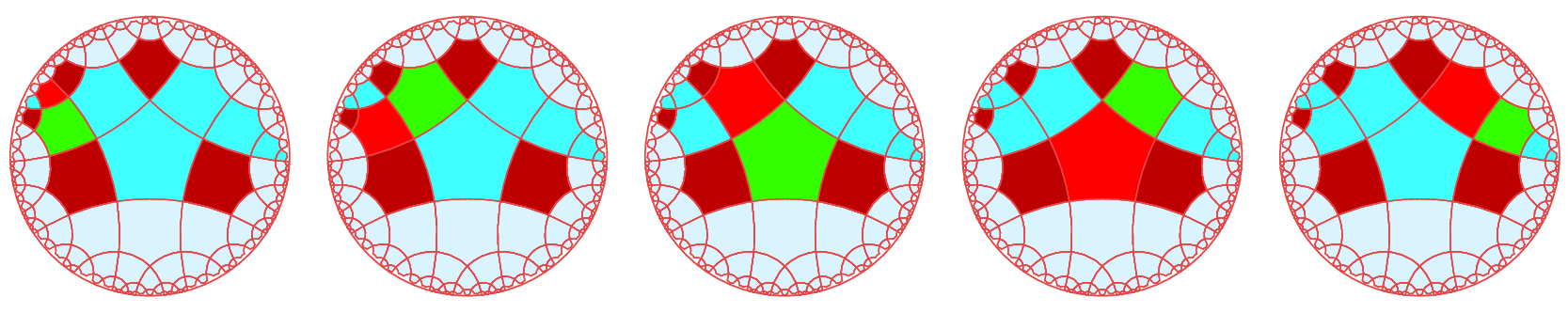}
\hfill}
\ligne{\hskip 5pt
\includegraphics[scale=0.4]{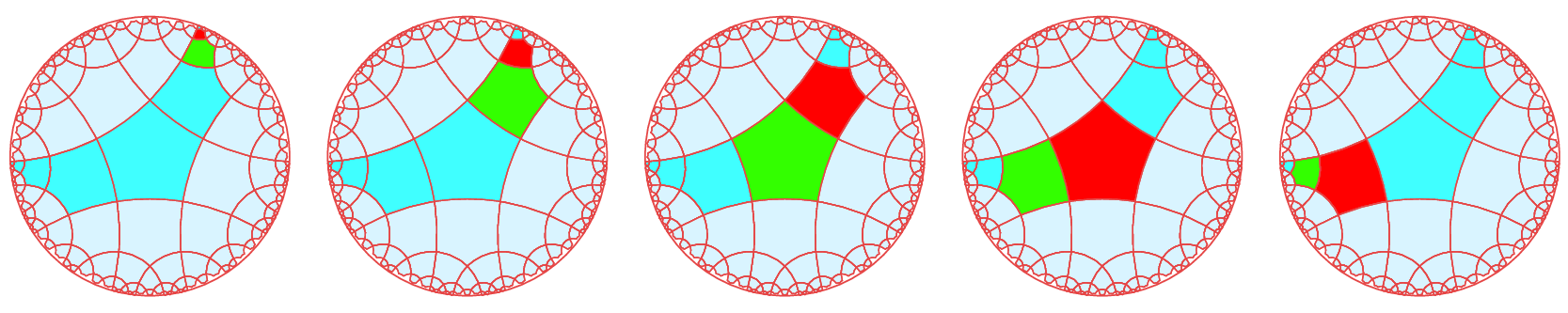}
\hfill}
%\vspace{-10pt}
\begin{fig}\label{hca54voies}
\leurre
Implantation des voies dans la pentagrille.\\
En haut, voie horizontale parcourue de gauche \`a droite.\\
En bas, voie verticale parcourue de haut en bas.
\end{fig}
\vspace{10pt}
}

\vtop{
\vspace{-5pt}
\ligne{\hskip 5pt
\includegraphics[scale=0.5]{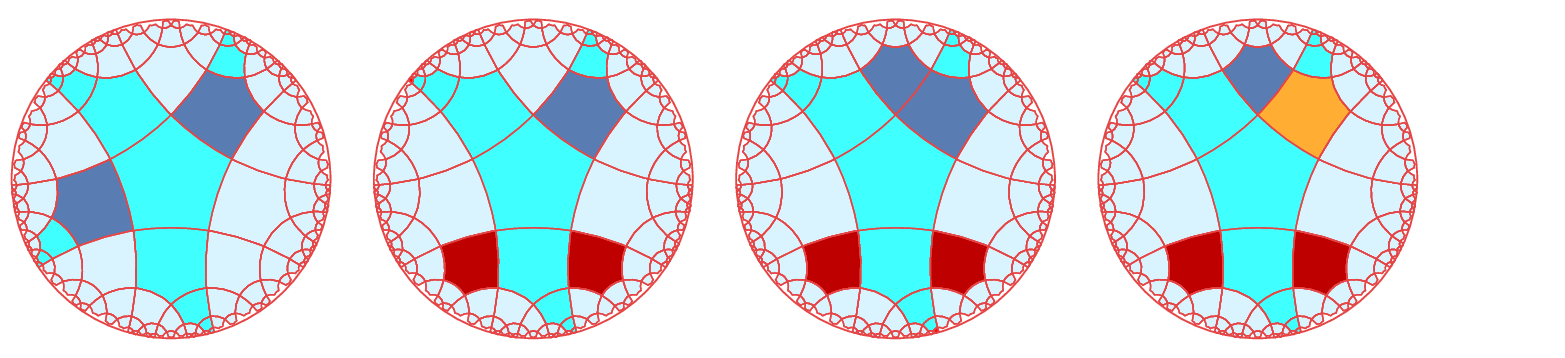}
\hfill}
%\vspace{-10pt}
\begin{fig}\label{hca54stables}
\leurre
Implantation du croisement et des aiguillages dans la pentagrille.\\
De gauche \`a droite~: le croisement, le fixe, le m\'emorisant et la bascule.
\end{fig}
\vspace{10pt}
}

   In this implementation, as well as in that of~\cite{fhmmTCS}, the locomotive is
implemented as two contiguous cell with different colours, which allows the implemented
vehicle to find the direction of its motion along the tracks. The colours were chosen as green
and red, green pointing at the front and red at the rear. The direction is then obvious.
These are the elements which allowed us to prove the following result.

\begin{thm}\label{univpenta} {\rm (Margenstern, Song),} {\it cf.}{\rm\cite{mmsyPPL}}
$-$ There is a planar cellular automaton on the pentagrid with $9$-states
which is weakly universal and rotation invariant.
\end{thm}

   By planar, we mean that the trajectory of the cells of the cellular automaton which at 
some point change their state is a planar structure which contains infinitely many cycles which
cannot reduced to a $1D$-structure. This is in particular the case of the units which constitute
the registers of the register machine implemented by the railway circuit.

\subsubsection{The new scenario}
\label{newscenar}

   In this paper, we take benefit of various improvements which I brought in the construction
of weakly universal cellular automaton constructed in other contexts: in the heptagrid,
another grid of the hyperbolic plane, in the hyperbolic $3D$-space and in the tiling
$\{13,3\}$ of the hyperbolic plane, that latter automaton having two states only,
see~\cite{mmbook3} for details.

   Our implementation follows the same general simulation as the one described in 
Subsection~\ref{railway}. In particular, Figure~\ref{hca54element} is still meaningful in
this new setting.

\vtop{
\vspace{-5pt}
\ligne{\hfill
\includegraphics[scale=0.5]{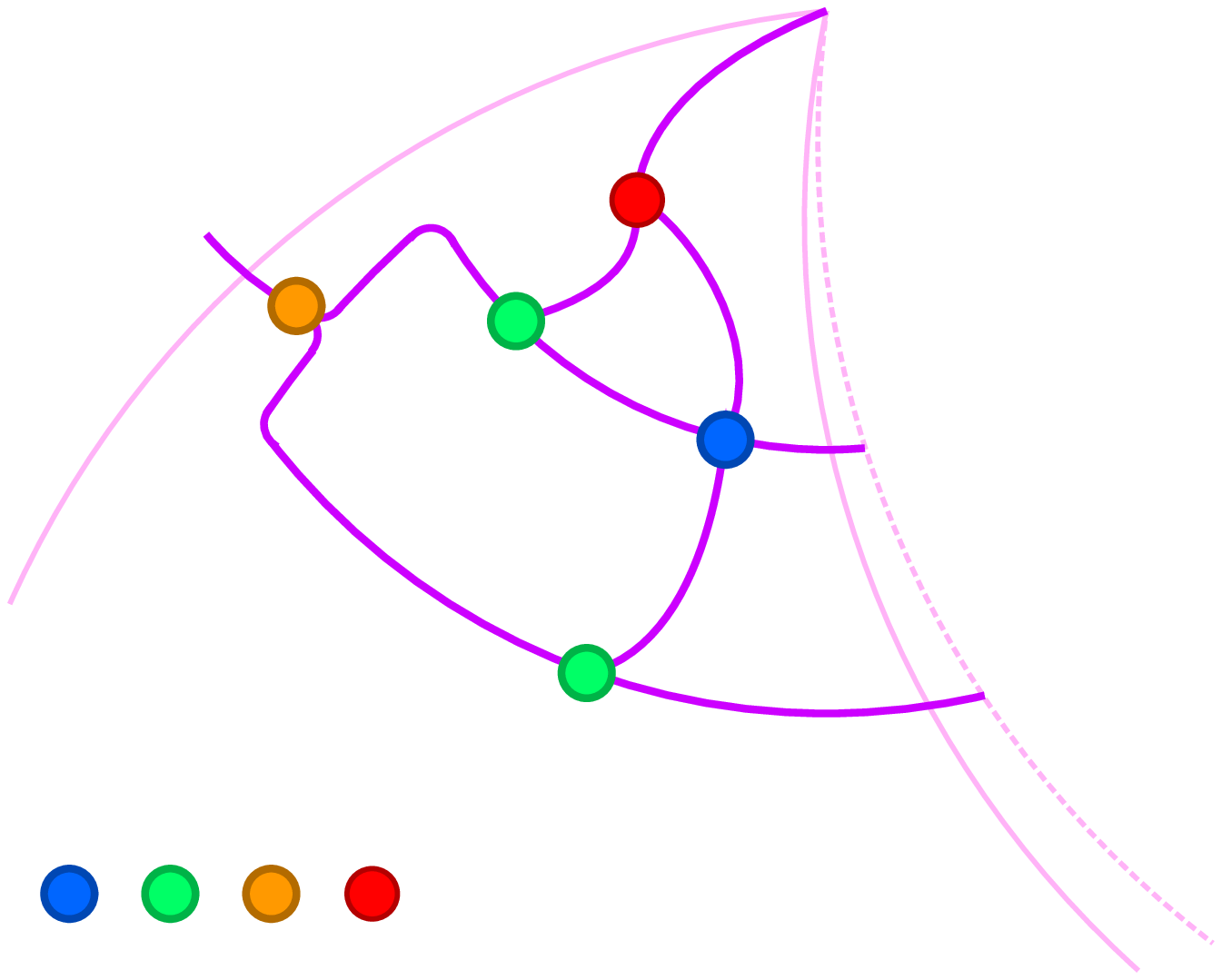}
\hfill}
%\vspace{-10pt}
\begin{fig}\label{hca54element}
\leurre
Implantation de l'\'el\'ement de base dans la pentagrille.\\
Les disques de la ligne du bas symbolisent, de gauche \`a droite~: le croisement, le fixe,
le m\'emorisant et la bascule.
\end{fig}
\vspace{10pt}
}

   However, here, new features are introduced.

   The first change is that the tracks are one-way. In some sense this is closer to what
we can see for railways in real life, in particular for highspeed ones. This change entails
a big change in the switches and in the crossings. There is no change for the flip-flop which
was already a one-way structure from the very beginning as passive crossings are ruled out
for this kind of switches. For fixed switches it introduces a very small change:we keep the
structure for a passive crossing and for the active one, as the selected track is the same, it 
is enough to continue the active way without branching at the centre of the switch,
see Figure~\ref{newswitches}. In the same picture, we can see that the situation is different
for the memory switch. This time, as there are two possible crossings of the switch and as the 
selected track may change, we have two one-way switches: an active one and a passive one. 
At first glance, the active switch looks like a flip-flop and the passive switch looks like
a one-way fixed one. However, due to the working of the memory switch, we could say that the
active memory switch is passive while the passive memory switch is active. Indeed, 
during an active crossing, the selected track is not changed contrary to what happens in the
case of the flip-flop. Now, during a passive crossing, the switch looks at which track is
crossed: the selected or the non-selected one. If it is the non-selected one, then the selection
is changed and this change is also transferred to the active switch. Accordingly, there is
a connection between the active and the passive one-way memory switches.

\vtop{
\vspace{-5pt}
\ligne{\hfill
\includegraphics[scale=0.5]{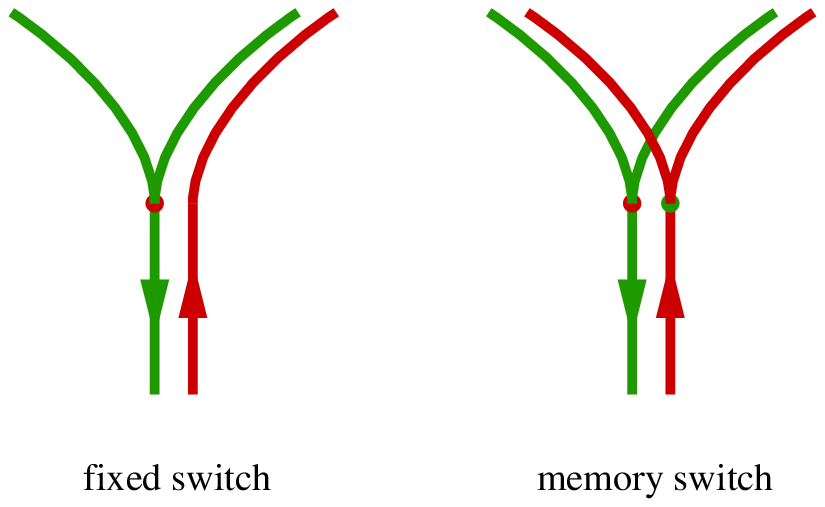}
\hfill}
\vspace{-20pt}
\begin{fig}\label{newswitches}
\leurre
The new switches for a one-way structured circuit: the fixed and the memory ones. Note that
the flip-flop remains the same as in Figure~{\rm\ref{switches}}.
\end{fig}
\vspace{10pt}
}

   Now, if we wish to significantly reduce the number of states, we also have to change the tracks 
themselves, as it appears that giving them the same colour as the background is better than
assigning a special colour to identify the tracks. The consequence is that we have to place 
{\it milestones} in order to do so. This was performed in previous works, see~\cite{mmbook3}.
But this is not enough: we also have to change the crossings. Contrary to what happens
in the $3D$-space where crossings can be replaced by bridges, which makes the situation
significantly easier, crossings cannot be avoided in the plane. 

   In~\cite{mmbook3} we indicate a solution which allowed me to build a weakly universal
cellular automaton in the hyperbolic plane with 2~states only. However, this was not performed
in the pentagrid nor in the heptagrid, but in the tiling $\{13,3\}$. This solution can be 
implemented here and this allowed me to reduce the number of states from~9 down to~5.
There is a slight improvement in the present solution which might allow us to reduce
the number of neighbours for a two-state weakly universal cellular automaton in the hyperbolic
plane.

   First, we look at a crossing of two one-way tracks. The main idea is that we organize
the crossing in view of a {\bf round-about}: an interference of road trafic in our railway 
circuit. We may notice that the locomotive arriving either from~{\bf A} or~{\bf B} in
Figure~\ref{rondpointsimple} has to turn right at the second pattern it meets on its way.
So that it is enough to devise a pattern which allows to count from~1 up to~2{} in some
way. In the two-states world of the weakly universal cellular automaton on $\{13,3\}$ 
described in~\cite{mmbook3}, there were four patterns: a first one appears when the locomotive
arrives at the round-about. At this point, a second locomotive is appended to the arriving one.
At the second pattern, one locomotive is removed and so, as a single locomotive arrives
through the round-about at the third pattern, then it knows that it has to turn right.
Here, this scheme is slightly changed as follows. At the first pattern it meets the locomotive,
which, arriving in a state~{\tt S}, is changed into~{\tt T}. At the second pattern, as a 
locomotive in the state~{\tt T} arrives, it is sent on the track which leaves the round-about. 
This is why three patterns are needed in Figure~\ref{rondpointsimple}.

\vtop{
\vspace{-5pt}
\ligne{\hfill
\includegraphics[scale=0.5]{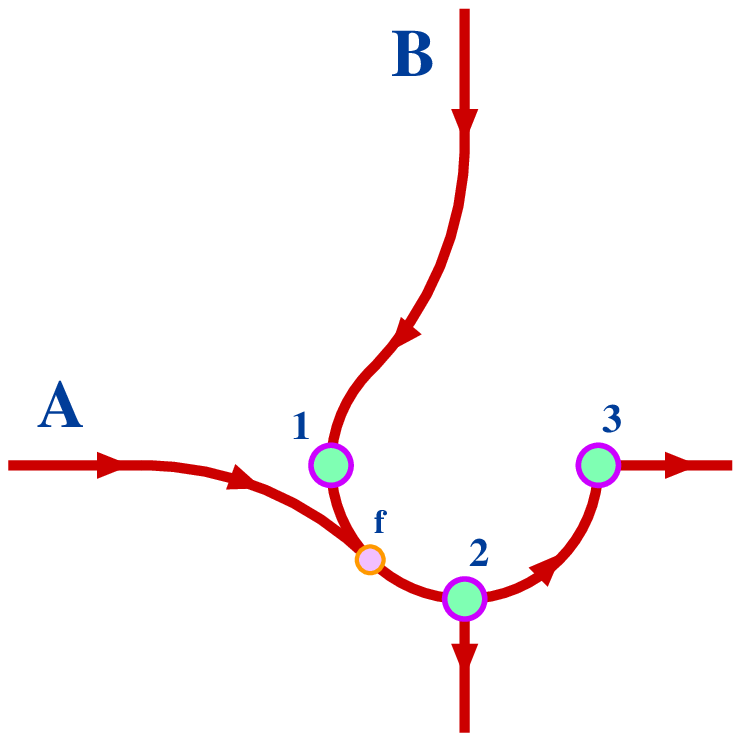}
\hfill}
\vspace{-20pt}
\begin{fig}\label{rondpointsimple}
\leurre
The new crossing: the one-way tracks from{\bf A} and {\bf B} intersect. We have a three-quarters 
round-about. 
The small disc at~{\bf\small f} represents a fixed switch. Discs~{\bf 1}, {\bf 2} and~{\bf 3}
represent the pattern which dispatches the motion of the locomotive on the appropriate way.
Patterns~{\bf 1} and~{\bf 3} are needed as explained in the description of the scenario.
\end{fig}
}

\vtop{
\vspace{-25pt}
\ligne{\hfill
\includegraphics[scale=0.5]{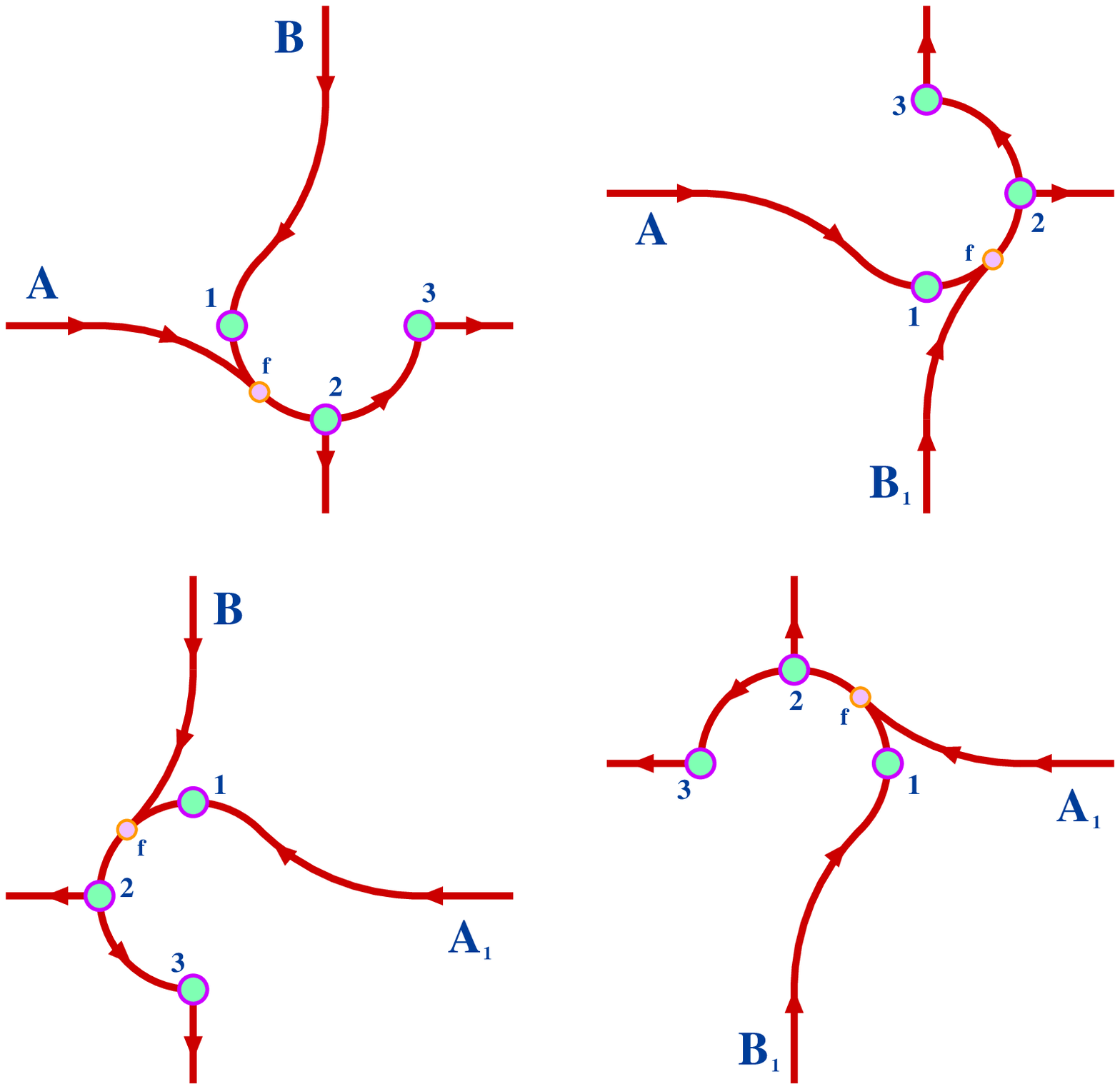}
\hfill}
\vspace{-10pt}
\begin{fig}\label{rondpointcomplet}
\leurre
The new crossing: four possible one-way track. Assembling them allows to perform a
two-way crossing. The notations are those of Figure~{\rm\ref{rondpointsimple}}. {\bf A}$_1$,
{\bf B}$_1$ go opposite to~{\bf A}, {\bf B}, respectively.
\end{fig}
\vspace{10pt}
}

   Figure~\ref{rondpointcomplet} shows us how to assemble four one-way round-abouts in order
to perform a true crossing fro two intersection two-ways tracks.

   Now that we have seen the scenario to implement the circuit, we have to precisely look
at how to implement it with five states only. We turn now to this question.

\subsubsection{Implementing the new scenario}
\label{scenar}
\def\YY{\hbox{\tt Y}{}}
   As already announced, we need to use five states only. The first state is the quiescent state
which we denote by \WW. Remember it is defined by the following rule: if a cell~$c$ and all
its neighbours are in the state~\WW, then at the next top of the clock, the cell~$c$ remains
in the state~\WW. We shall also call~\WW{} the blank and we also shall say that a cell in~\WW{}
is white. The other states are~\BB, \GG, \RR{} and~\YY. The cells in these states are said
to be blue, green, red or yellow, respectively. The state~\BB{} is mainly used for the 
milestones which delimit the tracks. The state~\GG{} is the basic state of the locomotive.
In crossings, the locomotive turns to~\RR. The state~\YY{} is used for special markings in
the passive memory switch. The states~\BB, \GG{} and~\RR{} are also used for marking in the 
crossings and in the other switches. The state~\GG{} also appears in the milestones for the tracks.
 
  Checking the new scenario requires to first study the implementation of the tracks. We have to
define verticals and horizontals.

\vskip 7pt
\noindent
\ligne{$\underline{\hbox{Vertical and horizontal tracks}}$}
\vskip 5pt

  The verticals are easy to define, they correspond to branches of the Fibonacci tree. In fact
they are finite sequences of cells for which a side lies on a fixed line. The line can be changed
inside the sequence as illustrated by Figure~\ref{voievertv} where a green locomotive goes on 
two verticals, one in a bottom-up run, the other in a top-down one.
%\section{Automates cellulaires universels dans le plan hyperbolique}
%\label{univplan}

%\section{Automates cellulaires universels dans l'espace hyperbolique $3D$}

\vtop{
\vspace{-5pt}
\ligne{\hfill
\includegraphics[scale=0.5]{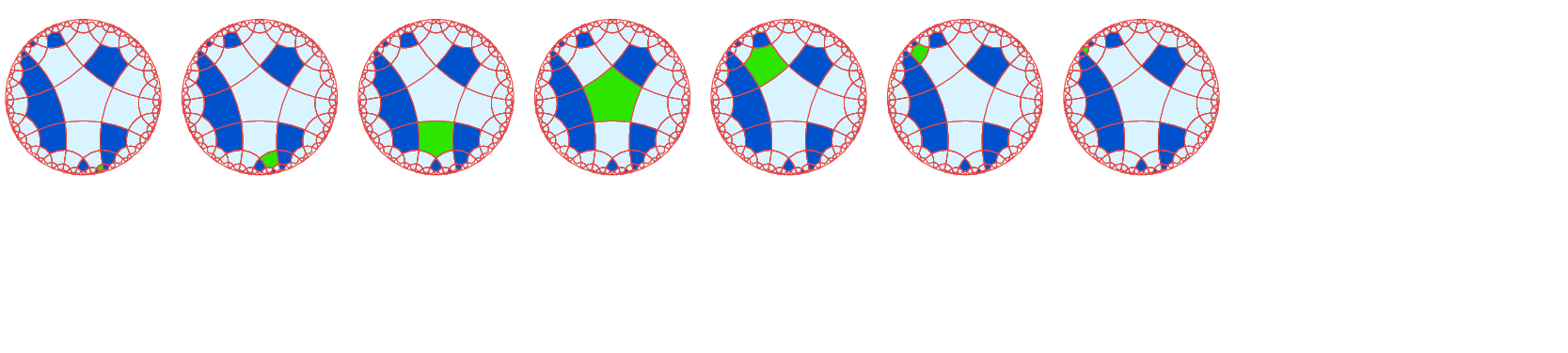}
\hfill}
\vspace{-40pt}
\ligne{\hfill
\includegraphics[scale=0.5]{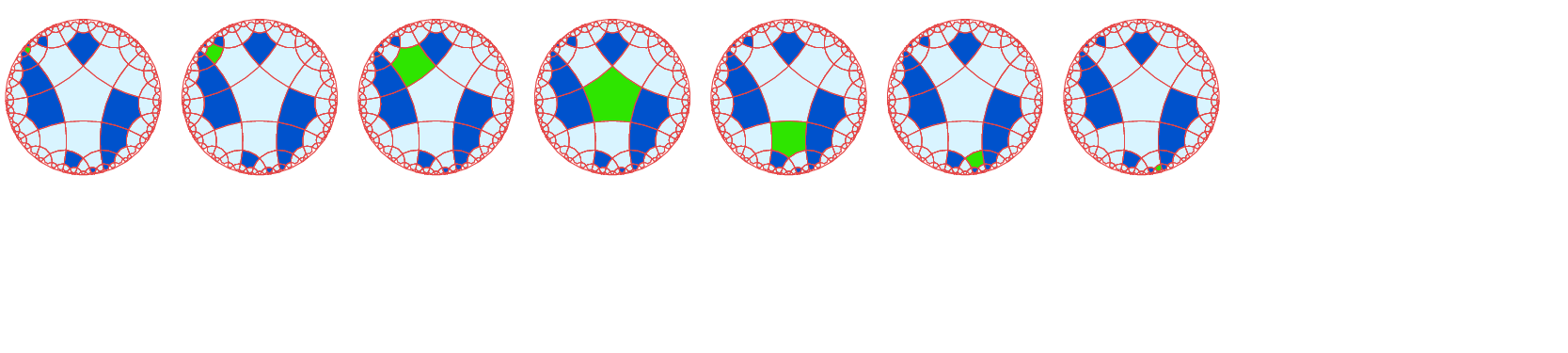}
\hfill}
\vspace{-40pt}
\begin{fig}\label{voievertv}
\leurre
A vertical with a green locomotive, 
First row: top-down traversal; second row: bottom-up traversal.
\end{fig}
\vspace{10pt}
}

   The structure of an element of the track is simple. Assume that the locomotive leaves the
cell through its side~1. Then, the milestones are always on sides~2 and~5. It enters either
through sides~3 or~4. Both cases occur as illustrated in Figures~\ref{voievertv}
and~\ref{voievertr}. Now, if we consider
the standard numbering of the sides, then the place of the milestones depends on the direction
of the motion. Assume that in the central cell, side~3 is the horizontal side. Then, for the
cell 1(1), cell~1 of sector~1, the milestones are on the sides~3 and~5{} in a bottom-up motion
while they are on the sides~2 and~5{} in the top-down one. The leftmost picture of
Figure~\ref{elemvoie} illustrates the milestones as just defined. 

\vtop{
\vspace{-5pt}
\ligne{\hfill
\includegraphics[scale=0.5]{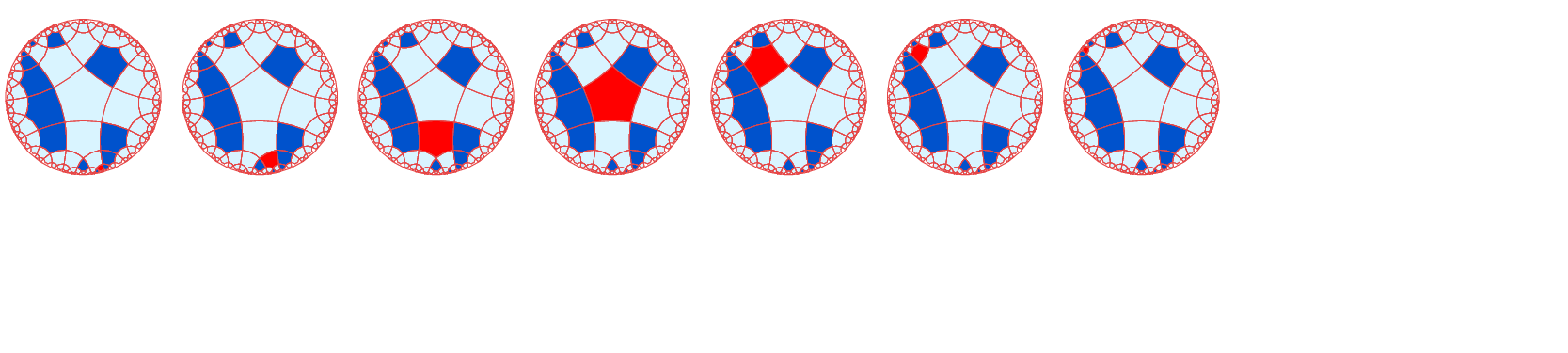}
\hfill}
\vspace{-40pt}
\ligne{\hfill
\includegraphics[scale=0.5]{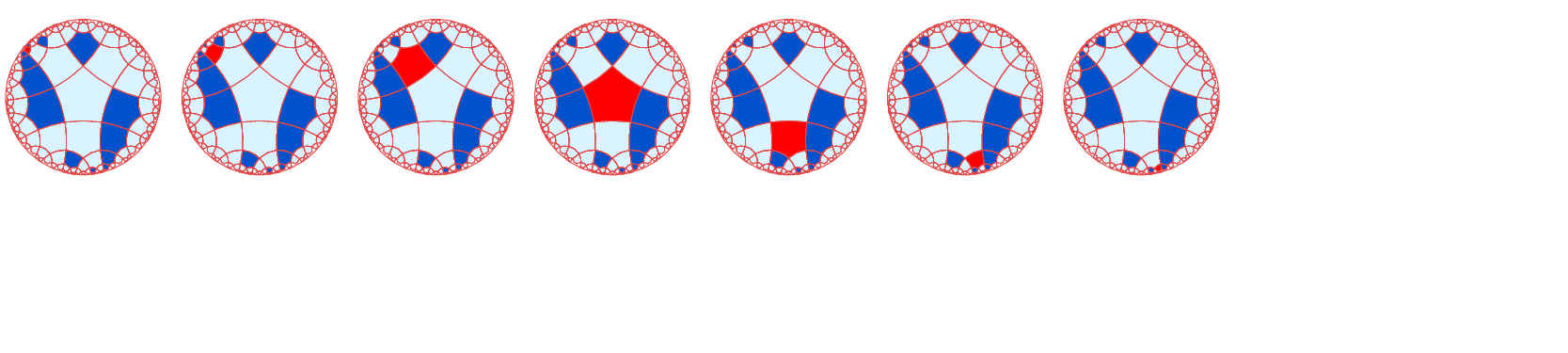}
\hfill}
\vspace{-45pt}
\begin{fig}\label{voievertr}
\leurre
A vertical with a red locomotive, 
First row: top-down traversal; second row: bottom-up traversal.
\end{fig}
\vspace{10pt}
}

\vtop{
\vspace{5pt}
\ligne{\hfill
\includegraphics[scale=0.4]{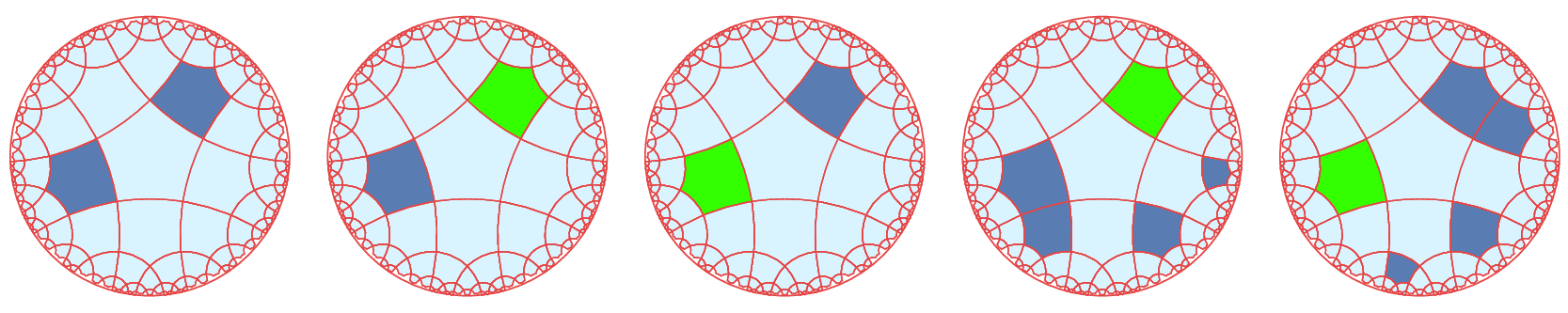}
\hfill}
\vspace{-5pt}
\begin{fig}\label{elemvoie}
\leurre
The elements of the tracks. \vskip 0pt
Leftmost picture: the standard element. Second and third pictures from left: the element
which allows to perform sharp turns. Fourth and fifth pictures, illustration for a sharp
turn.
\end{fig}
\vspace{10pt}
}

The other pictures of Figure~\ref{elemvoie} illustrate two other patterns for the tracks and
their importance. The second and third picture of the figure illustrates an element of the
track which allows the locomotive, either green or red, to perform a sharp turn: this means
that the locomotive enters through a side and exits through a contigous one. Such a turn
is absolutely needed in the pentagrid in order to have cycles in the trajectory of the locomotive.
If such a possibility would not be allowed, the locomotive would run to infinity without 
returning to any tile it had already visited.

   The patterns illustrated by the second pictures allow us to perform a sharp turns as it
is illustrated by the last two pictures. A blue milestone is replaced by a green one. There are
two possiblities and each of them is used for a direction of the motion. Rules are devised
in such a way that the green milestone prevents a backward motion of the locomotive. Indeed,
if in the fourth picture, the green milestone is replaced by a blue one, then a locomotive going
from sector~3 into sector~4 would go on its way in sector~4 but a copy of it would return
to tile~1 of sector~3. This backward motion is prevented by the green milestone when it
is in tile~1 of sector~5. Now, the sharp turn is possible as the exit of the locomotive
does not occur through the expected side~1 but through a side which is contiguous to the
one through which the locomotive entered the cell. 

\vtop{
\vspace{0pt}
\ligne{\hfill
\includegraphics[scale=0.5]{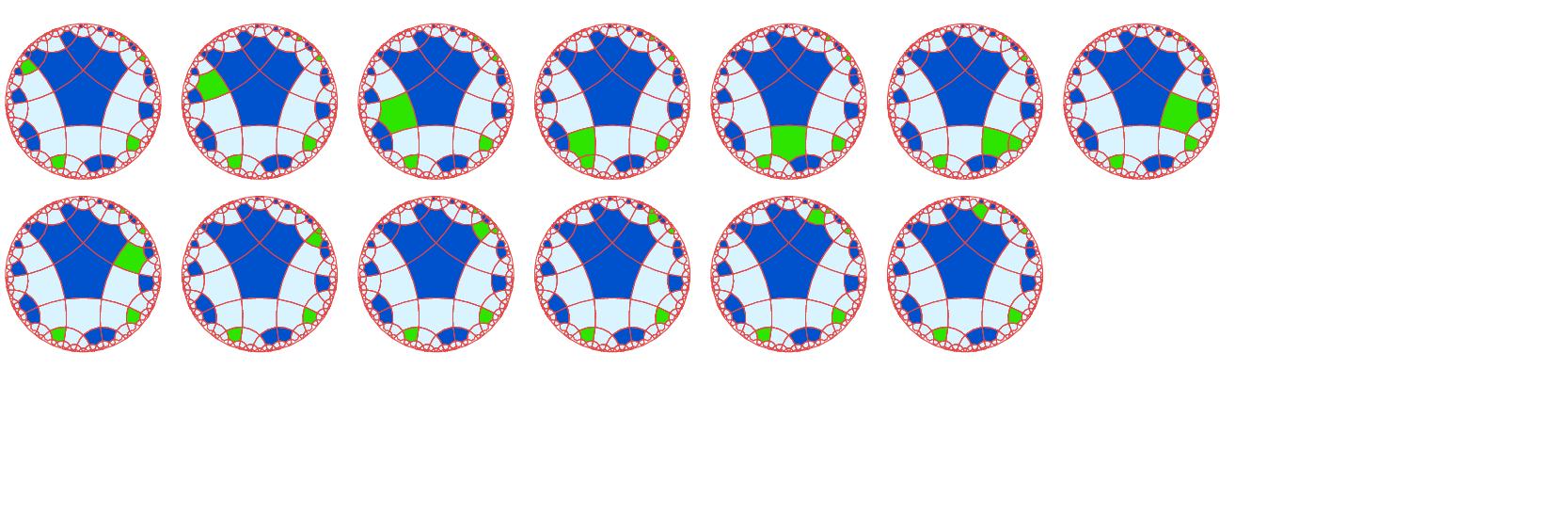}
\hfill}
\vspace{-40pt}
\ligne{\hfill
\includegraphics[scale=0.5]{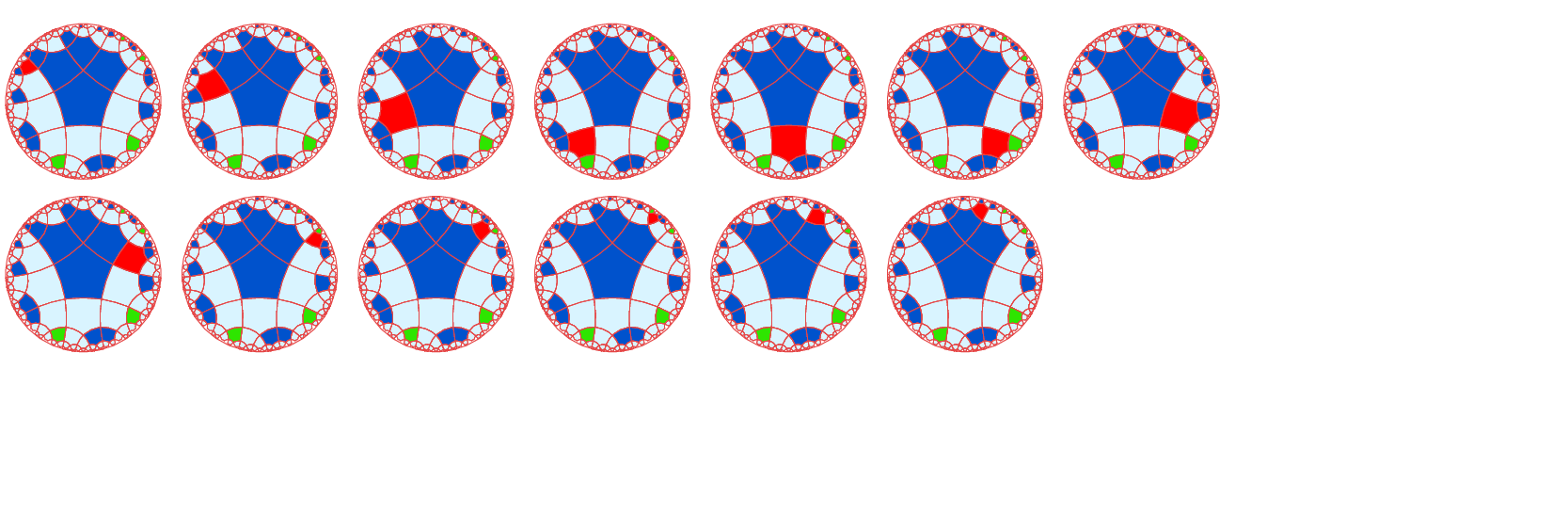}
\hfill}
\vspace{-40pt}
\ligne{\hfill
\includegraphics[scale=0.5]{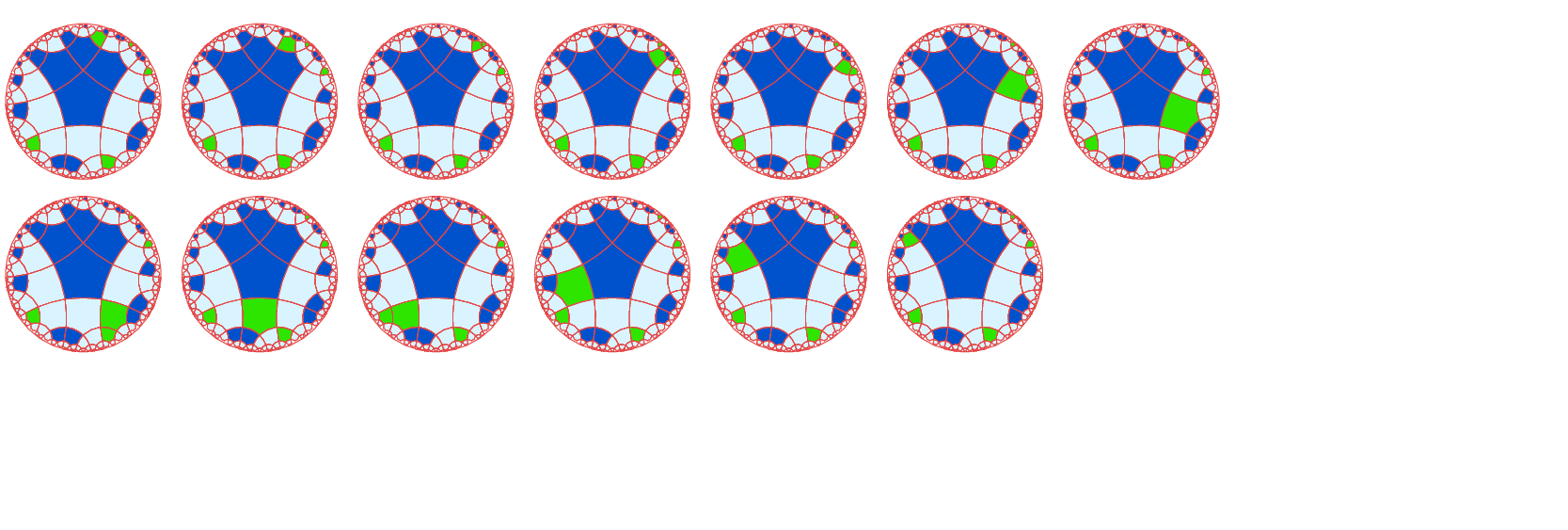}
\hfill}
\vspace{-40pt}
\ligne{\hfill
\includegraphics[scale=0.5]{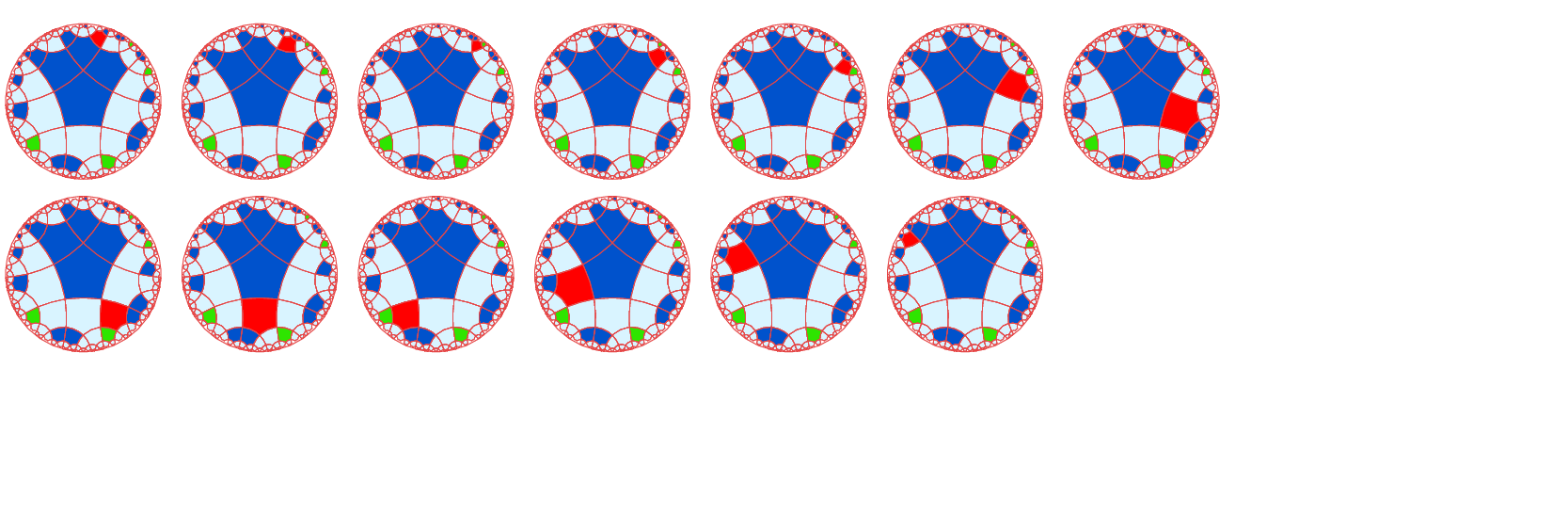}
\hfill}
\vspace{-45pt}
\begin{fig}\label{voiehoriz}
\leurre
A horizontal with a green locomotive, first four rows and then, a red one: the last four rows.
For each locomotive, right-left and left-right runs.
\end{fig}
\vspace{10pt}
}

Figure~\ref{zoomhor} allows us to establish that any horizontal can be run by the locomotive
with the elements indicated in Figure~\ref{elemvoie}. The horizontals follow the levels
which are defined in~\cite{mmbook1}: they belong to bigger and bigger Fibonacci trees whose
roots follow a vertical as defined above. Weprove that the track can go from a black node
on a level to the next one. There are four situations, depending on the father of the black node 
and the arrival to the black node, either from the upper level or from its son, on the lower
level. On Figure~\ref{zoomhor} three horizontals
are indicated by a thin coloured line which crosses the tiles, a move line, a green one and 
a red one. Note that contiguous tiles on the same horizontal do not share an edge, they only share
a vertex. In Figure~\ref{zoomhor} and in the next figures in this subsection, we shall
use the following numbering of the sectors: the sectors are counterclockwise numbered 
from~1 to~5 and sector~3 is defined by the pentagon which is below the horizontal side
of the central tile. In each sector~$s$ the tile is given by its number~$n$, we write
$n(s)$ if it is needed to indicate the sector.

Figure~\ref{zoomhor} indicates the motion from a black node~$\beta$ on a given level, the green one
in the figure, to the next one~$\gamma$ on the same level and in the considered direction. Both 
directions are illustrated in the figure in order to facilitate the checking. Now, a black node 
may have either a white or a black father and the track may arrive at~$\beta$
either from the upper level or from the lower one. This explains the four cases illustrated on 
each row of the figure. Note that in the motion from left to right, the track goes through
2(2), 0 and 2(5) when the father of~$\beta$ is black and it goes 
through 2(2), 0 and 2(1) when the father is white. In the motion from right to left, the
visited cells occur in the opposite order. In the figure, the central cell is the father
of~$\beta$ in the motion from right to left. From Figure~\ref{voiehoriz}, we can check that
a horizontal can be run by the locomotive, whether it is green or red.

\vtop{
\vspace{5pt}
\ligne{\hfill
\includegraphics[scale=0.4]{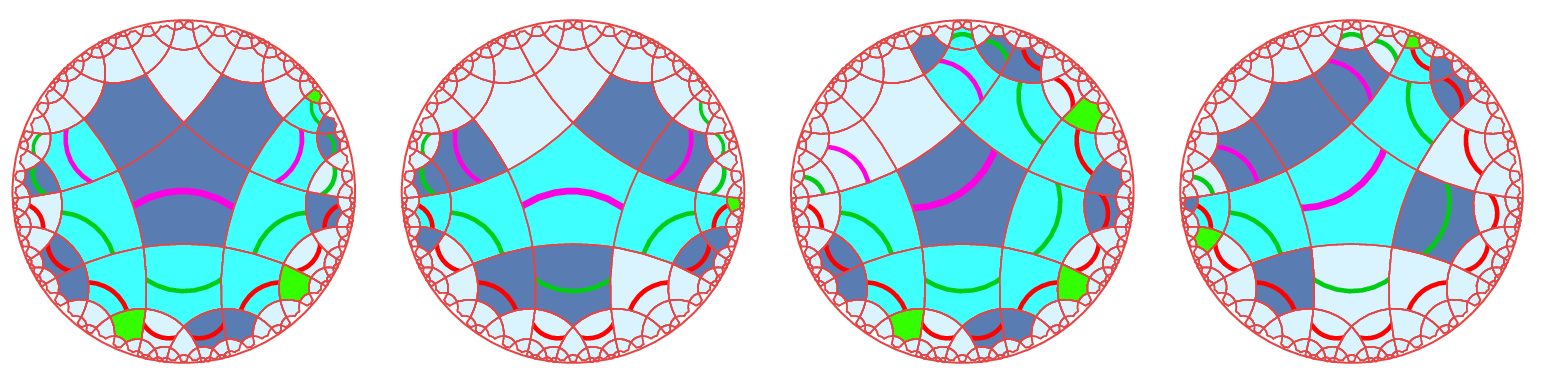}
\hfill}
\ligne{\hfill
\includegraphics[scale=0.4]{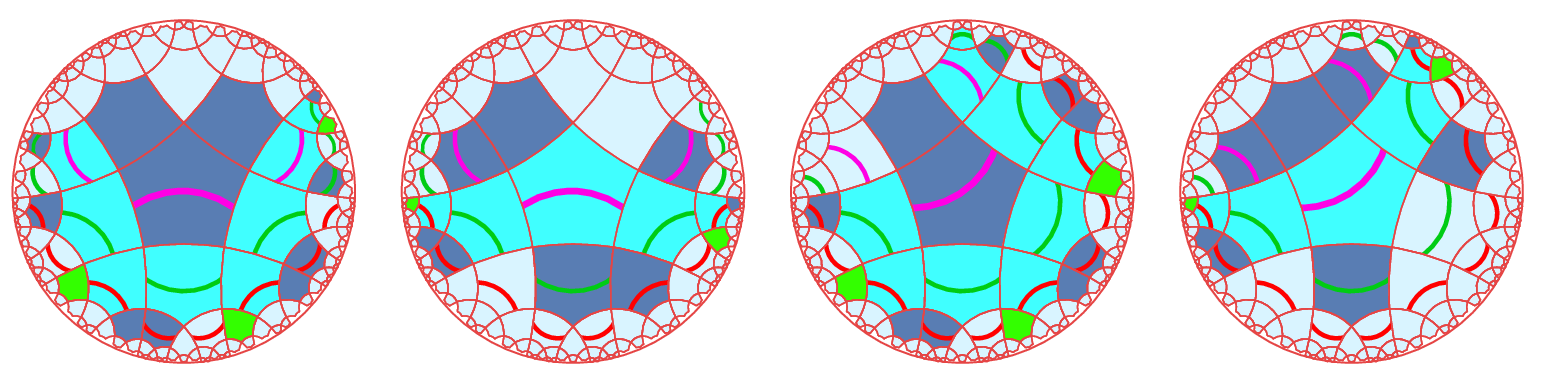}
\hfill}
\vspace{-5pt}
\begin{fig}\label{zoomhor}
\leurre
Closer look on a horizontal. The central cell is the father of~$\beta$ in the motion from left
to right.  Note the indication of the horizontals through the coloured lines. In white nodes
the line has a $\smile$~shape, in black nodes it is a $\frown$-shape.
\vskip 0pt
First row: from left to right; second row: from right to left.
\end{fig}
\vspace{10pt}
}

   Figure~\ref{voiehoriz} shows that the connection with a vertical, either up or down or to the
left or to the right raises no problem. The figure as well as Figures~\ref{voievertv}
and~\ref{voievertr}
also shows that such tracks can be run indifferently by a green or a red locomotive.

   It is the place to remark that as horizontal tracks run on three consecutive levels, a two-way
protion of the tracks require at least seven consecutive levels as we need at least one level to
separate the tracks run on each direction. A similar remark also holds for vertical lines.
A two-way section must be separated by several nodes on the same level at the level where the
distance between the supporting line is minimal: this distance must be positive, which guarantees 
that the lines are not secant. These constraints require much space for the implementation, but
in the hyperbolic plane, we are never short of space.

\vskip 7pt
\noindent
\ligne{$\underline{\hbox{Implementing the crossings}}$}
\vskip 5pt

   As indicated in Subsubsection~\ref{scenar}, the tracks are organized according to
what is depicted in Figures~\ref{rondpointsimple} and~\ref{rondpointcomplet}. From the
latter figure, it is enough to focus on the implementation of Figure~\ref{rondpointsimple}.
From the implementation of the tracks, we only have to look at the implementatation
of the patterns symbolically denoted as~{\bf 1}, {\bf 2}, {\bf 3} and~{\bf f} in the
figure. As {\bf f} is a fixed switch whose implementation is indicated a bit further,
we simply implement~{\bf 1} as the other patterns are strict copies of this one.
Figure~\ref{hca54crois} illustrates this implementation and the behaviour of the locomotive
when it crosses the pattern. We can see that the difference strongly depends on the colour of
the locomotive.

   From Figure~\ref{rondpointsimple}, the locomotive always arrives to the pattern from the same
cell, namely 4(4). Then the locomotive goes to the centre, cell~0. 

\vtop{
\vspace{-55pt}
\ligne{\hfill
\includegraphics[scale=0.6]{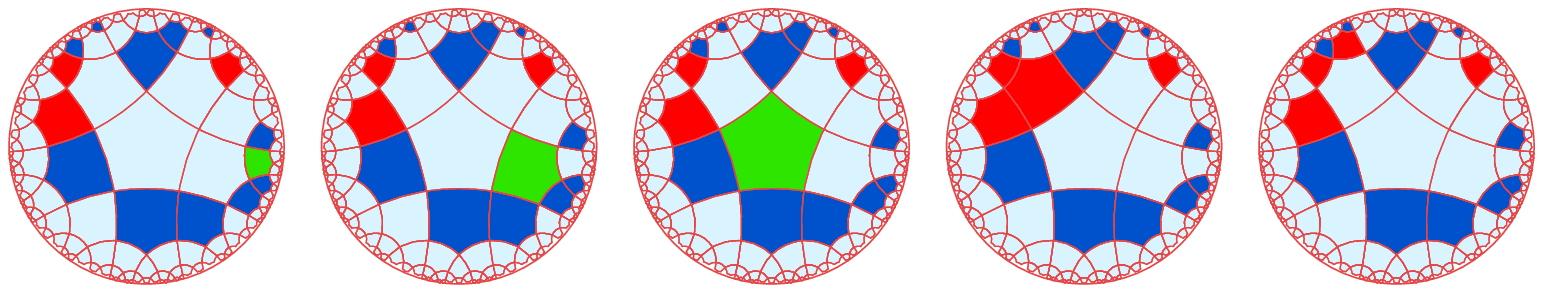}
\hfill}
\vspace{-50pt}
\ligne{\hfill
\includegraphics[scale=0.6]{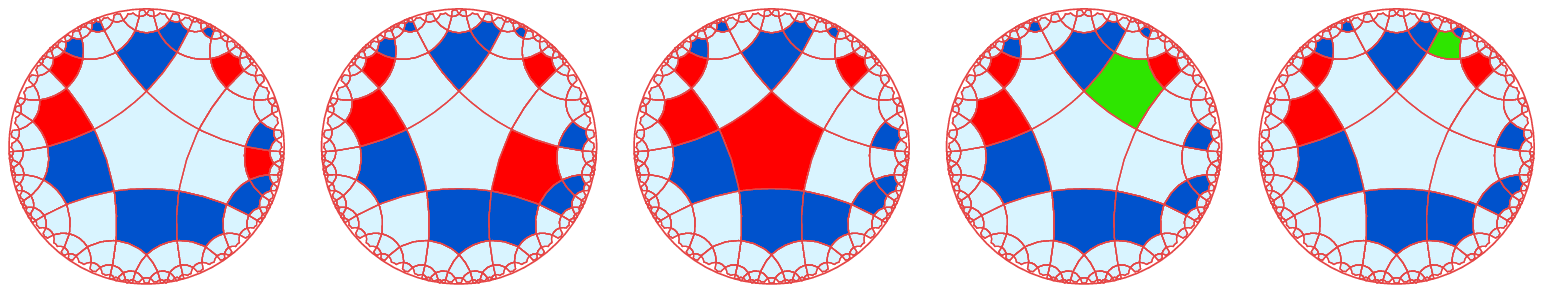}
\hfill}
\vspace{-5pt}
\begin{fig}\label{hca54crois}
\leurre
The key pattern of the crossings. Notice the difference of behaviour depending on the
colour of the locomotive.
\end{fig}
\vspace{10pt}
}

   When the locomotive is green, it goes
to cell 1(1) arriving there as a red cell, and then it goes out onto the round about towards
the next pattern, still as a red locomotive. When the locomotive which arrives at cell 4(4) is
red, it also goes to~0 but from there, it goes to 1(5) where it arrives as a green cell.
Indeed, cell 1(5) remains white until it sees a red cell through its side~1. Note that side~1
is clearly indentified thanks to the pattern of the neighbours of cell~1(5). It is the pattern
of an element of the tracks with a red milestone in place of the green one.

\vskip 7pt
\noindent
\ligne{$\underline{\hbox{Implementation of a fixed switch}}$}
\vskip 5pt

   The first two rows of Figure~\ref{hca54fix} illustrate the passive crossing of a fixed 
switch for a green locomotive while the last two rows of the same figure does the same for 
the red locomotive.

   With one-way tracks, we need a single kind of passive fixed switch. Indeed, there is no need
of an active fixed switch, as the locomotive never goes in the non-selected direction. The active
switch is reduced to the track which goes in the selected direction, as illustrated by
Figure~\ref{newswitches}.

\vtop{
\vspace{-5pt}
\ligne{\hfill
\includegraphics[scale=0.5]{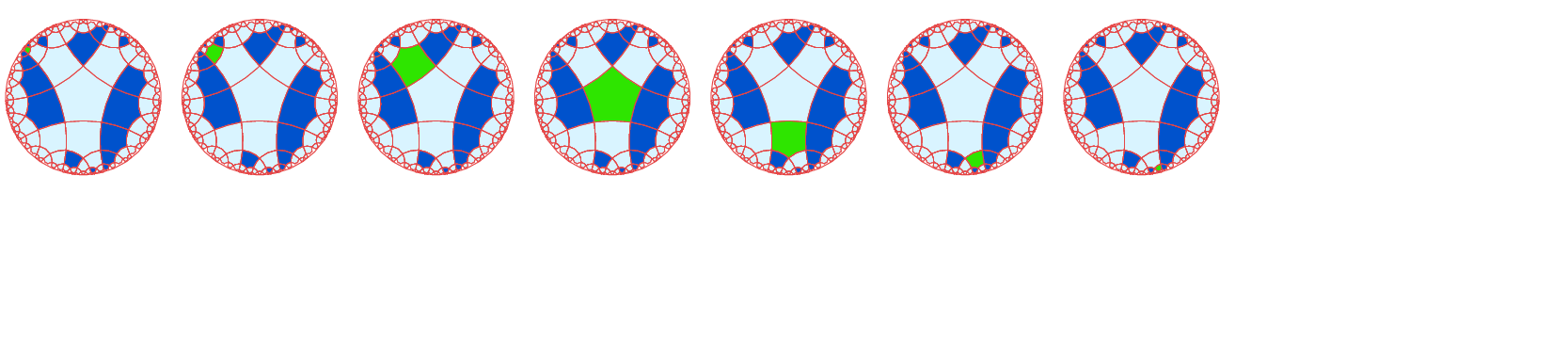}
\hfill}
\vspace{-45pt}
\ligne{\hfill
\includegraphics[scale=0.5]{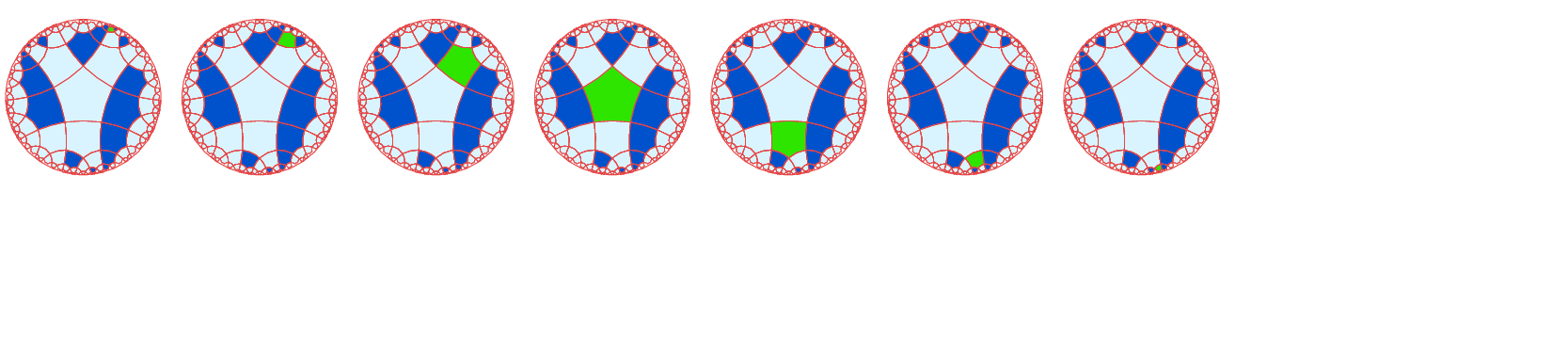}
\hfill}
\vspace{-45pt}
%\begin{fig}\label{hca54fixv}
%The fixed switch, passive crossing by the green locomotive.
%\end{fig}
%}
%
%\vtop{
%\vspace{-5pt}
\ligne{\hfill
\includegraphics[scale=0.5]{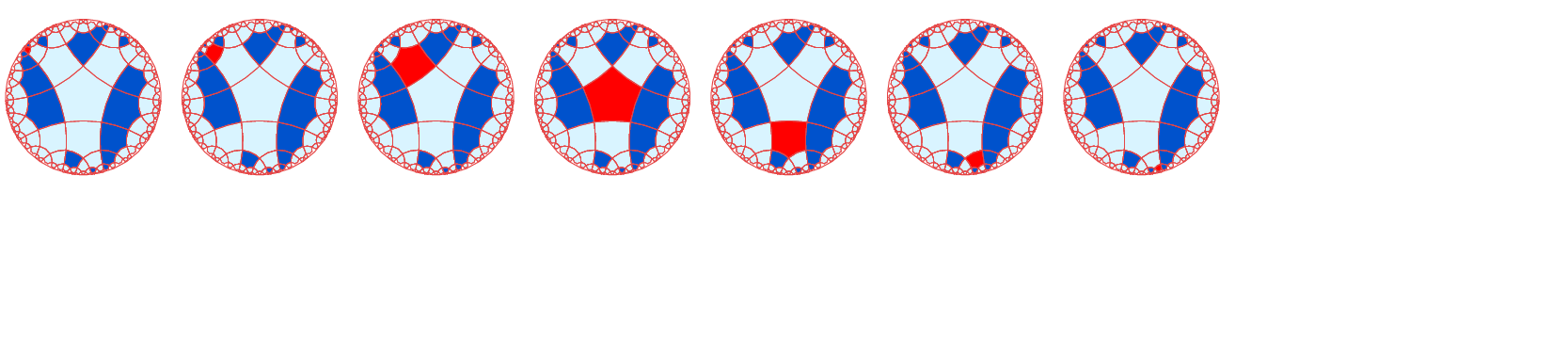}
\hfill}
\vspace{-45pt}
\ligne{\hfill
\includegraphics[scale=0.5]{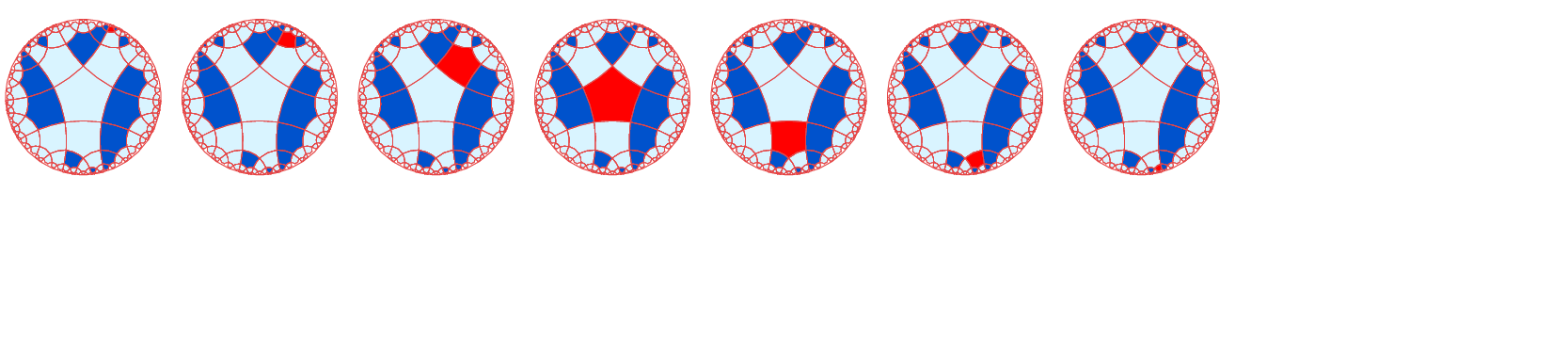}
\hfill}
\vspace{-45pt}
%\begin{fig}\label{hca54fixr}
\begin{fig}\label{hca54fix}
\leurre
The fixed switch: passive crossing by the green locomotive and then the red one.
\end{fig}
\vspace{10pt}
}

\vskip 7pt
\noindent
\ligne{$\underline{\hbox{Implementation of a flip-flop}}$}
\vskip 5pt

   Yp to now, we used four states only, those which were introduced with the tracks. It was not
difficult to implement the tracks and it was possible to implement the crossings using still
these states. With the flip-flop, we introduce the fifth state, \YY. Figure~\ref{hca54basc}
illustrates the configuration of a flip-flop and its active crossing by a locomotive.

\vtop{
\vspace{-5pt}
\ligne{\hfill
\includegraphics[scale=0.5]{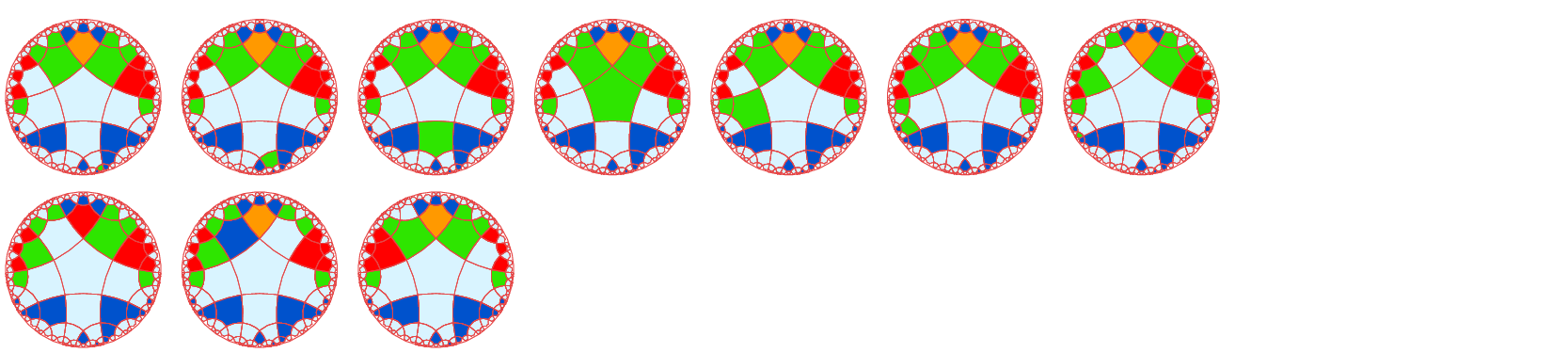}
\hfill}
\vspace{-5pt}
\ligne{\hfill
\includegraphics[scale=0.5]{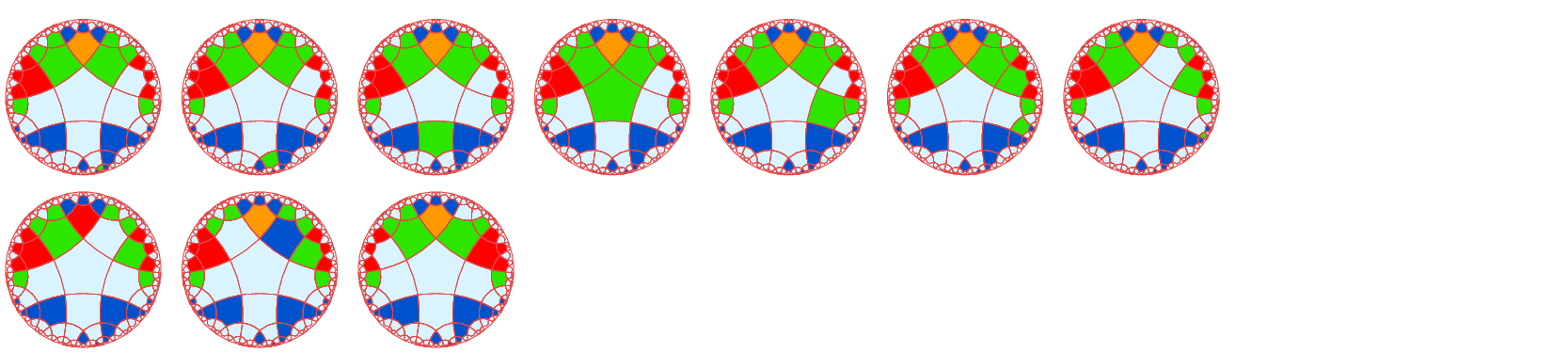}
\hfill}
\vspace{-5pt}
\begin{fig}\label{hca54basc}
\leurre
The flip-flop. 
\end{fig}
\vspace{10pt}
}

   We can imagine that the locomotive crosses a second time the flip-flop, turning it back to the
position it had before the first crossing. We remark that the change of the selected track occurs
some time after the locomotive left the switch. In fact, the white cell at~2(2), it has 
three red neighbours, detects the passage of the locomotive just before the latter leaves the
switch. Then, it passes the information to the cell~2(1) through 1(1). This makes 2(1) to flash,
turning from~~\YY{} to~\RR{} and then turning back to~\YY. This flash makes both~1(1) and~1(5)
to change their states in a way which triggers~2(2) and~2(5) to change their states. When
the cell 2(2) is red, a symmetric process occurs.

\vskip 7pt
\noindent
\ligne{$\underline{\hbox{Implementation of a memory switch}}$}
\vskip 5pt

    Now, we arrive to the most difficult situation. We have to implement two switches with
a connection between them. As already noticed in Subsection~\ref{scenar}, the active switch
has a passive behaviour when crossed by the locomotive and the passive switch has an active
behaviour when the locomotive takes the non-selected track. This action of the passive switch
triggers the change of selection in the active switch: hence we have to organize the connection
from the passive witch to the active one. The patternsof these switches, when the locomotive 
is not present, is illustrated by Figure~\ref{hca54memostab}.

\vtop{
\vspace{-5pt}
\ligne{\hfill
\includegraphics[scale=0.4]{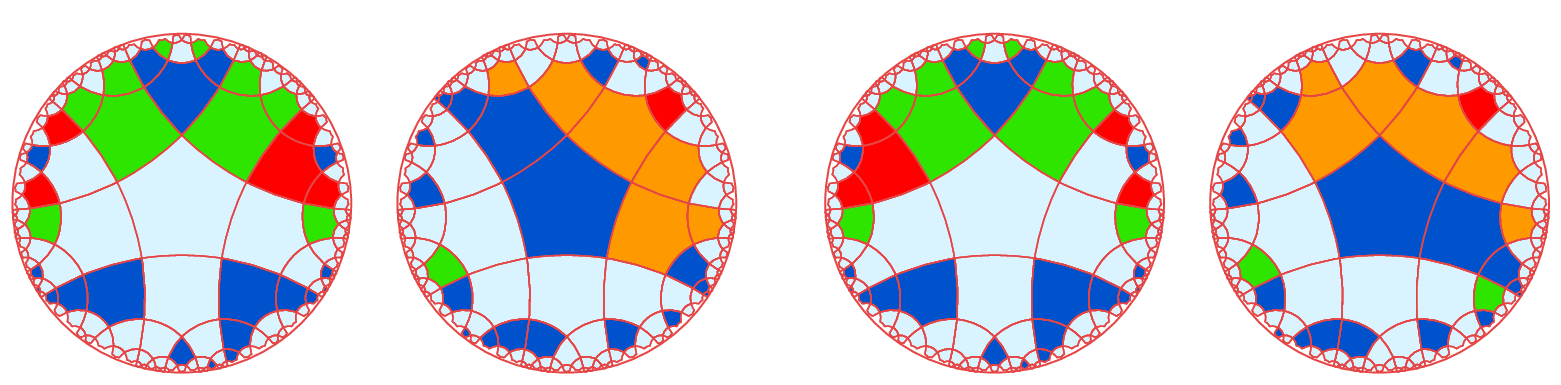}
\hfill}
\vspace{-5pt}
\begin{fig}\label{hca54memostab}
\leurre
The stable configuration of the active and passive memory switches. To left, the switches
selected the right-hand side track. To right, they selected the left-hand side track. 
\end{fig}
\vspace{5pt}
}

   Figure~\ref{hca54memoactif} illustrates the crossing of the switch by the locomotive.
Due to its way of working, the memory switch has two basic positions according to which is
the selected track. We say that the left-, right-hand side switch selects the left-, right-hand
side track respectively. We can check on the figure that
at the switch remains unchanged adter the traversal
of the locomotive. Note that the switch remains unchanged after the locomotive crossed the
switch.

\vtop{
\vspace{-5pt}
\ligne{\hfill
\includegraphics[scale=0.5]{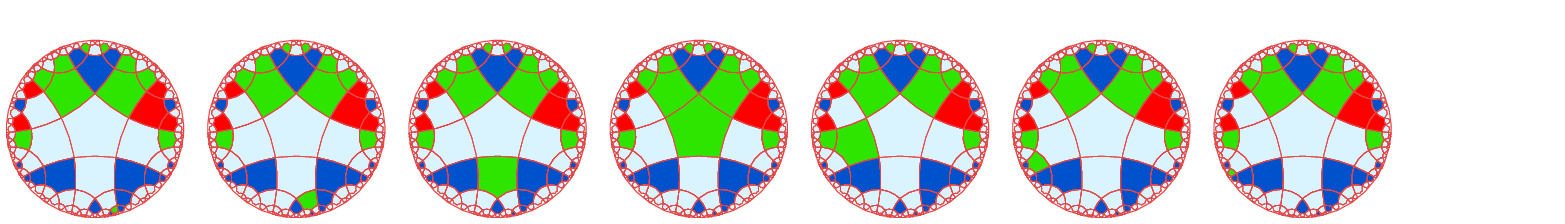}
\hfill}
\vspace{-5pt}
\ligne{\hfill
\includegraphics[scale=0.5]{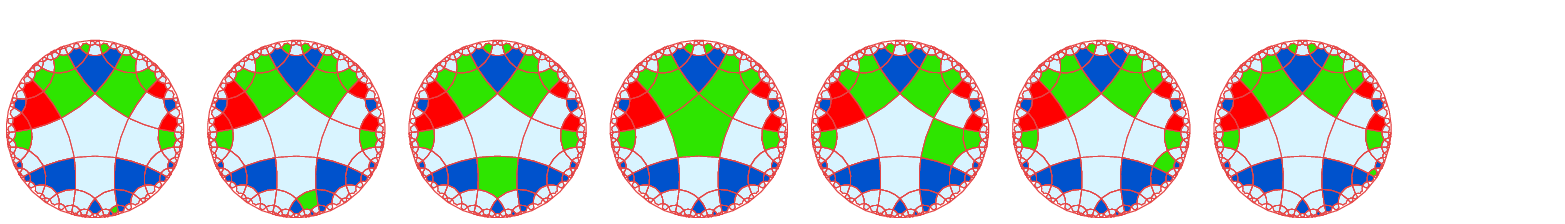}
\hfill}
\vspace{-5pt}
\begin{fig}\label{hca54memoactif}
\leurre
The active memory switch. The two versions of the switch: above, left-hand side switch,
below, right-hand side one. 
\end{fig}
\vspace{10pt}
}

   We can see that the pattern of the active memory switch looks like that of the flip-flop.
The difference is restricted to the cell~2(1) and its neighbours. In the flip-flop,
the cell~2(1) is in~\YY{} and three consecutive neighbours ar in~\BB: cells~5, 6 and~7 of
sector~1. Now, in the active memory switch, the cell~2(1) is in~\BB{} and the cell~6(1) 
is white. Moreover, the cells~15(1) and~18(1) are in~\GG. We shall see the role of these green
cells in a while. 

   Before, we look at the crossing of the passive memory switch. We have four situations
as the switch has two positions and as for each position, the locomotive may arrive either
through the selected track or through the non-selected one.

\vtop{
\vspace{-45pt}
\ligne{\hfill
\includegraphics[scale=0.5]{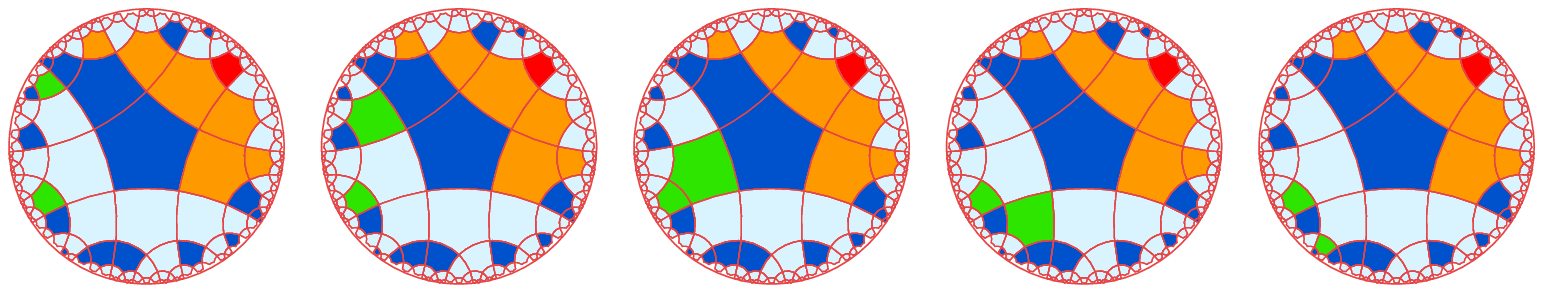}
\hfill}
\vspace{-5pt}
\ligne{\hfill
\includegraphics[scale=0.5]{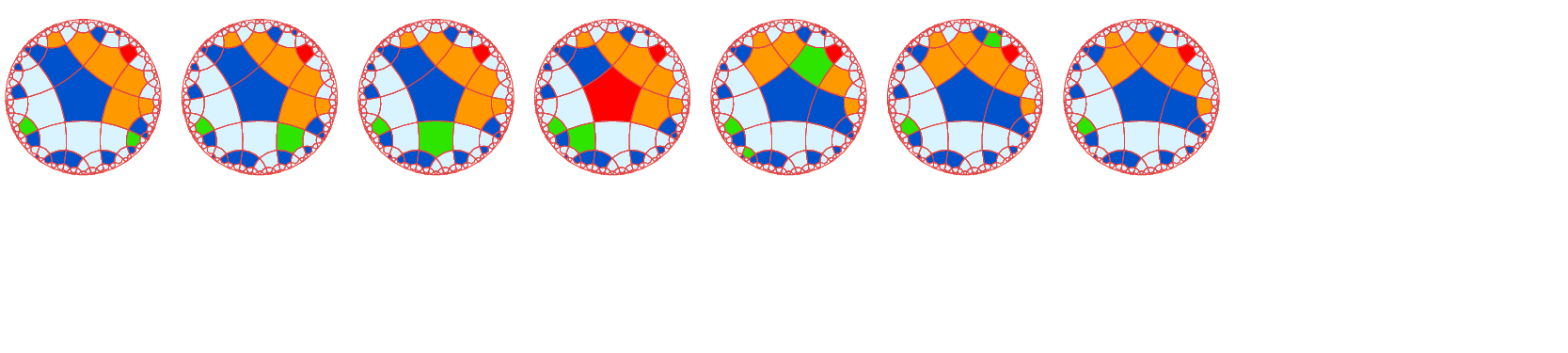}
\hfill}
\vspace{-85pt}
\ligne{\hfill
\includegraphics[scale=0.5]{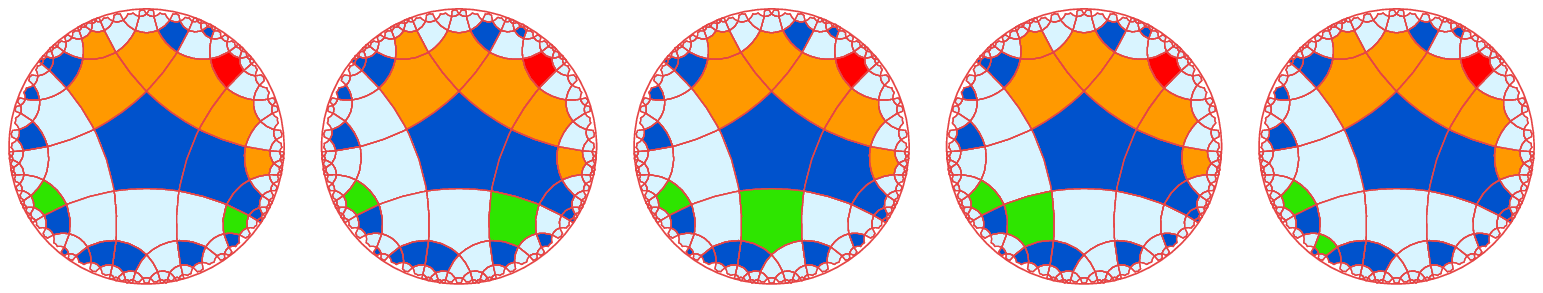}
\hfill}
\vspace{-5pt}
\ligne{\hfill
\includegraphics[scale=0.5]{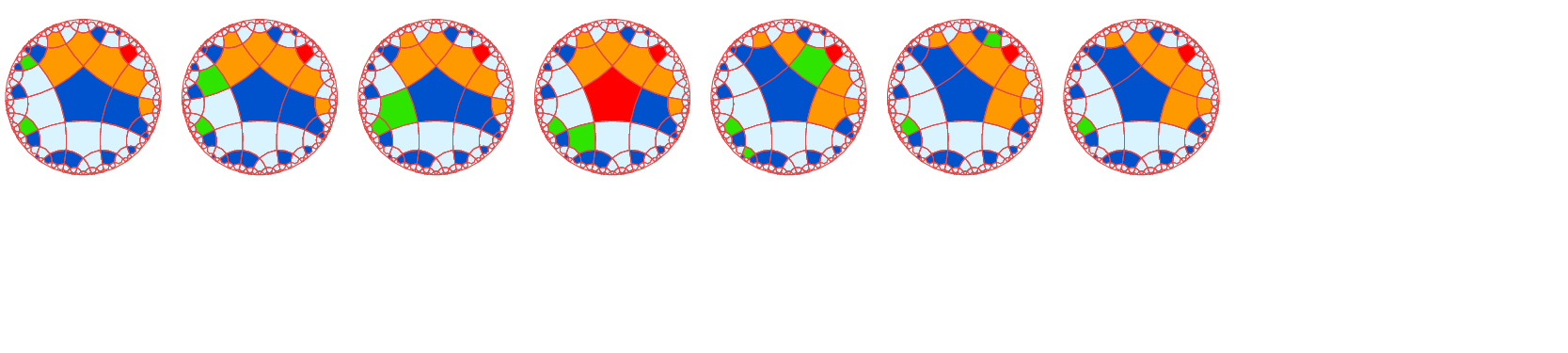}
\hfill}
\vspace{-45pt}
\begin{fig}\label{hca54memopassif}
\leurre
The passive memory switch. The two versions of the switch and the two possible
crossings: above, first two rows, left-hand side switch; below, last two rows, right-hand side one. 
\end{fig}
\vspace{10pt}
}

   As is clear on Figure~\ref{hca54memostab}, the pattern of the passive memory switch
is very different from the active one. Although a part of it is taken from a fixed switch
whose centre would be at the cell 2(3), the surrounding of this cell and the passive tracks
is very specific.

   Figure~\ref{hca54memopassif} illustrates all the four cases of crossing the switch. In the 
figure, it can be checked that when the locomotive arrives through the selected track, nothing
is changed outside the tracks themselves. It can also be checked that when the locomotive
crosses the non-selected track, this changes the selection according to the definition
of the memory switch: the non-selected track becomes the new sekected track.

   Now, this information has to be transferred to the active memory switch. This is performed
by the pattern of the passive memory switch. The crossing through the non-selected track is
detected by the cell~1(5)i which is in contact with the central cell. That latter one has 
a \BB-neighbour on the side of the selected track and a \YY-one on the side of the non-selected 
track. When the locomotive becomes a neighbour of the central cell, it abuts the cell on the
side of the \YY-neighbour or of the \BB-one. This allows the central to know whether the
locomotive has run through the selected track or through the non-selected one. Accordingly,
if the run goes through the non-selected one, the central cell flashes: it turns from~\BB{}
to~\RR{} and then turns back to~\BB. Now, this flash makes the cells~1(1) and~1(4) to take
the opposite colour, from~\BB{} to~\YY{} or from~\YY{} to~\BB: this changes the signalization
of the non-selected track. But the cell~1(5) also can see the flash of the central cell.
This makes the cell~1(5) to also flash: it turns to~\GG{} and then turns back to~\YY. Now,
the cells~5(1) and~11(5) are milestones for the cell~4(5) which, accordingly, appears to
be a possible element of a track. Consequently, the flash of~1(5) creates a second locomotive
which can go along the tarck whose starting point is cell~4(5). It is enough to define a track
going to the active memory switch to make the needed connection between the two parts of the
memory switch.

\vtop{
\vspace{-5pt}
\ligne{\hfill
\includegraphics[scale=0.4]{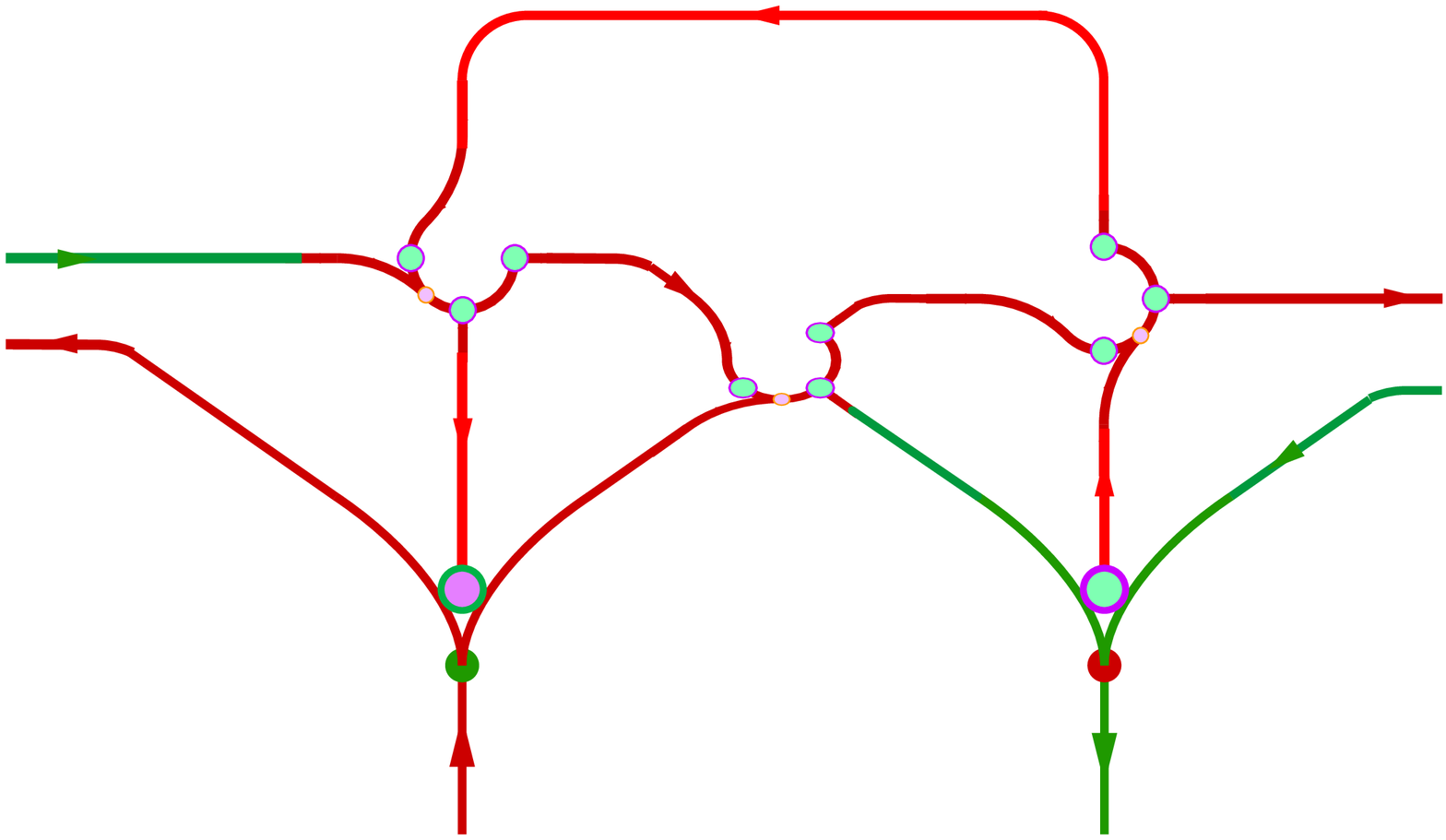}
\hfill}
\vspace{-5pt}
\begin{fig}\label{hca54memo}
\leurre
The organisation of the memory switch.
\end{fig}
}

   Figure~\ref{hca54memo} illustrates the global setting for implementing the memory switch
with its tow parts, the passive and the active one and the connection between them. Now, the path
whose starting point is the cell~4(5) in the passive switch, see Figure~\ref{hca54memostab},
goes to the active switch as indicated in Figure~\ref{hca54memo} and it arrives at the
cell~6(1) of the active switch, see Figure~\ref{hca54memostab}. 

\vtop{
\vspace{-45pt}
\ligne{\hfill
\includegraphics[scale=0.45]{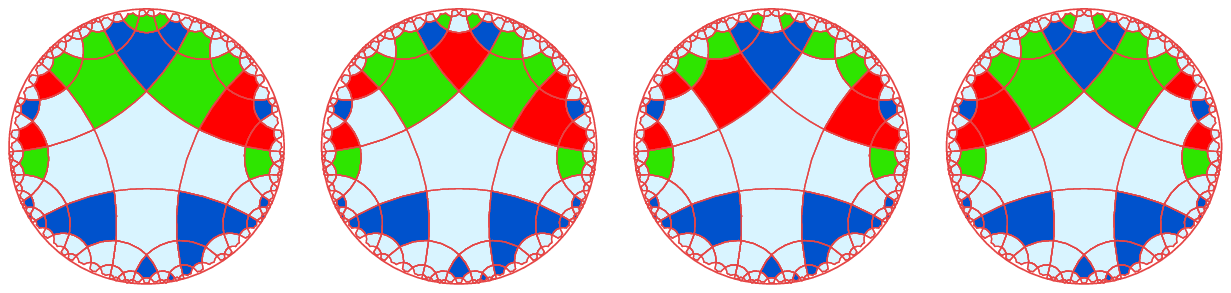}\hskip -180pt
\includegraphics[scale=0.45]{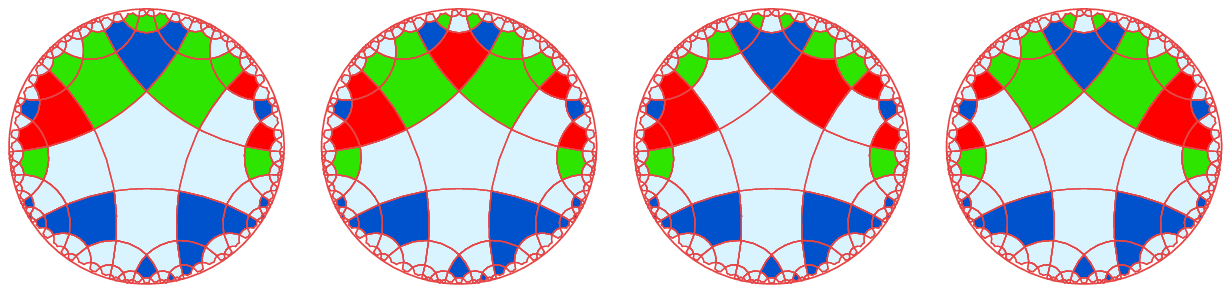}
\hfill}
\vspace{-5pt}
\begin{fig}\label{hca54memotrigger}
\leurre
The change of selection in the active memory switch triggered by the arrival of the second locomotive
at the cell~$6(1)$. To left the case when the left-hand side track is selected, to right, when it
is the case for the right-hand side track. The presence of the locomotive in~$6(1)$ can be seen 
in the first picture of each series.
\end{fig}
\vspace{5pt}
}

The cells 15(1) and~17(1)
allow to make the second locomotive sent from the passive switch go to the cell~16(1) from
where it is driven to the cell~6(1). Now, when the cell~2(1) can see the second locomotive,
it flashes, turning to~\RR{} and then back to~\BB, which makes the cells~1(1) and~1(5)
trigger the signal to the cells~2(2) and~2(5) which then take the opposite colour.

   Figure~\ref{hca54memotrigger} illustrates the situation when the second locomotive arriving
at the cell~6(1) triggers the change of selection. As can be seen in the figure, the second
locmotive vanishes just after it arrived at~6(1). This arrival makes the cell 2(1) flash and then
the same mechanism as seen for the flip-flop apply: the situation for cells~2(2) and~2(5) is exactly
the same.

   And so, outside the locomotive which yields the simulation of the computation in this model,
call it the {\bf main locomotive}, from time to time a second locomotive appears for a while in 
order to transmit the appropriate signal to an active memory switch. It is important to notice that 
the motion of this second locomotive does not interfer with the motion of the main one. Indeed,
although the track from a passive memory cell to its corresponding active one is very long, the 
distance between whole switches is much larger. In any case it can be made much larger: this can 
easily be seen on Figure~\ref{hca54element}. Also note that the second locomotive may sometimes
be red. Indeed, as indicated by Figure~\ref{hca54memo}, the second locomotive travels through two
crossings.

\subsection{The rules}
\label{rules}

   In order to prove the existence of the cellular automaton whose working was
descrived in the previous sections, we have to implement its rules. Here, we display 
all of them, but we list them in several groups according to the presentation of the previous
implementation. 
We present the rules according to the same order as we presented the implementation in
Subsection~\ref{scenar}

\subsubsection{Rules for the tracks}

   First, we have the rules for the motion of the locomotive on the tracks which are displayed
in Table~\ref{rultrack}. The rules are {\bf rotation invariant}.

   This means that if we perform a circular permutation on the neighbours of a cell, this
does not change the new state. Accordingly, in the tables we give, the rules are rotationally
invariant: a given rule~$\rho$ has four over rules with the same current state and the same 
new state, the states of the neighbours being a circular permutation of those indicated by~$\rho$.

   The rules are presented in the same order as established while checking them during the 
construction by a computer program.

   As explained in the caption of the table, the rules are divided into {\bf conservative} ones
and {\bf motion} ones. This is a general feature: at each top of the clock, the locomotive is
in one cell. There are at most two locomotives at a given time so that, at most 12~cells
may change from one time to the next one. Accordingly, the rest of the configuration must not
change. This is the role of the {\bf conservative} rules: they keep the configuration invariant,
the locomotives being supposed to be absent. The {\bf motion} rules control the whole simulation.
They define the changes of states but also they sometime do not change the state: either they
simply witness the passage of a locomotive, this is typically the case for milestones,
either this lack of change is part of the process: this is what we observe in the management of
the switches. 

\newcount\numero\numero=0
\def\arule #1 #2 #3 #4 #5 #6 #7 #8 {%
\footnotesize\tt
\hbox{%
\ifnum #1=0
  {\hbox to 20pt{\hfill}}
  \else
  {\global\advance \numero by 1
   \hbox to 20pt {\hfill\the\numero\hskip 5pt}}
  \fi
\hbox to 57pt {$\underline{\hbox{\tt #2}}$#3#4#5#6#7$\underline{\hbox{\tt #8}}$\hfill}
}
}

%nombre d'états : 5  W  B  G  R  Y  
\newdimen\largoo\largoo=92pt

\setbox110=\vtop{\leftskip 0pt\parindent 0pt\hsize=\largoo
\baselineskip 10pt
\ligne{\hfill
\arule 1  W W W W W W W 
\hfill}
\ligne{\hfill
\arule 1  W B W W W W W 
\hfill}
\ligne{\hfill
\arule 1  W G W W W W W 
\hfill}
\ligne{\hfill
\arule 1  W W W W B G W 
\hfill}
\ligne{\hfill
\arule 1  W W W W G B W 
\hfill}
\ligne{\hfill
\arule 1  W W W W B B W 
\hfill}
\ligne{\hfill
\arule 1  W G G W W W W 
\hfill}
\ligne{\hfill
\arule 1  W R G W W W W 
\hfill}
\ligne{\hfill
\arule 1  B W W W W W B 
\hfill}
\ligne{\hfill
\arule 1  G W W W W W G 
\hfill}
\ligne{\hfill
\arule 1  B W W W W B B 
\hfill}
\ligne{\hfill
\arule 1  B W W W W G B 
\hfill}
\ligne{\hfill
\arule 1  G W W W W G G 
\hfill}
\ligne{\hfill
\arule 1  B W W W B B B 
\hfill}
\ligne{\hfill
\arule 1  B W W W B G B 
\hfill}
}

\setbox112=\vtop{\leftskip 0pt\parindent 0pt\hsize=\largoo
\baselineskip 10pt
\ligne{\hfill
\arule 1  B W W W G B B 
\hfill}
\ligne{\hfill
\arule 1  B B B W B W B 
\hfill}
\ligne{\hfill
\arule 1  B B B W W B B 
\hfill}
\ligne{\hfill
\arule 1  B W W B W B B 
\hfill}
\ligne{\hfill
\arule 1  B W W B G B B 
\hfill}
\ligne{\hfill
\arule 1  B B B W B G B 
\hfill}
\ligne{\hfill
\arule 1  B B G W W B B 
\hfill}
\ligne{\hfill
\arule 1  B B W G W B B 
\hfill}
\ligne{\hfill
\arule 1  B B W W G B B 
\hfill}
\ligne{\hfill
\arule 1  B B B W B R B 
\hfill}
\ligne{\hfill
\arule 1  B B B G W B B 
\hfill}
\ligne{\hfill
\arule 1  B B B R W B B 
\hfill}
\ligne{\hfill
\arule 1  B B R W W B B 
\hfill}
\ligne{\hfill
\arule 1  B B W R W B B 
\hfill}
\ligne{\hfill
\arule 1  B B W W R B B 
\hfill}
}

\setbox114=\vtop{\leftskip 0pt\parindent 0pt\hsize=\largoo
\baselineskip 10pt
\ligne{\hfill
\arule 1  W W W B W B W 
\hfill}
\ligne{\hfill
\arule 1  W W W B W G W 
\hfill}
\ligne{\hfill
\arule 1  W W W G W B W 
\hfill}
\ligne{\hfill
\arule 1  W W G B W B G 
\hfill}
\ligne{\hfill
\arule 1  W W G B W G G 
\hfill}
\ligne{\hfill
\arule 1  W G W B W B G 
\hfill}
\ligne{\hfill
\arule 1  W G W B W G W 
\hfill}
\ligne{\hfill
\arule 1  W G W G W B G 
\hfill}
\ligne{\hfill
\arule 1  W R W B W B R 
\hfill}
\ligne{\hfill
\arule 1  W W R B W B R 
\hfill}
\ligne{\hfill
\arule 1  W W R B W G R 
\hfill}
\ligne{\hfill
\arule 1  W R W G W B R 
\hfill}
}

\setbox116=\vtop{\leftskip 0pt\parindent 0pt\hsize=\largoo
\baselineskip 10pt
\ligne{\hfill
\arule 1  W W W B G B W 
\hfill}
\ligne{\hfill
\arule 1  W W W B G G W 
\hfill}
\ligne{\hfill
\arule 1  W W W B R B W 
\hfill}
\ligne{\hfill
\arule 1  G W W B W B W 
\hfill}
\ligne{\hfill
\arule 1  G W W B W G W 
\hfill}
\ligne{\hfill
\arule 1  G W W G W B W 
\hfill}
\ligne{\hfill
\arule 1  R W W B W B W 
\hfill}
\ligne{\hfill
\arule 1  R W W B W G W 
\hfill}
\ligne{\hfill
\arule 1  R W W G W B W 
\hfill}
}

\vtop{
\vspace{-10pt}
\begin{tab}\label{rultrack}
\leurre
The rules for the tracks. The first two columns are conservative rules.
The last column contains the motion rules.
\end{tab}
\vspace{-10pt}
\ligne{\hfill\box110\hfill\box112\hfill\box114\hfill\box116\hfill}
}

\vskip 10pt
The motion rules controlling the motion of the locomotive, whatever its colour, 
define the change
of state which translates in terms of the cellular automaton the fact that the locomotive goes
from one cell of the track to the next one, see rules~31 up to~51. But the locomotive must 
not go to a white neighbour of the track which is not an element of the track. This property is 
shared by the motion rules which are conservative: this is the case for rules~43-43, it corresponds 
to the exit of the locomotive from the considered element of the track. It is also the case for 
rule~37 which prevents the entrance into an element where one of the two milestones 
is green: 
the entrance is allowed from the side of the blue milestone, rule~35, and forbidden from 
the side of the green milestone, it is the role for rule~37.

\subsubsection{Rules for the crossing and the fixed switch}

   Table~\ref{rulcrossfix} provides additional rules for the crossing and for the fixed switch.

\setbox110=\vtop{\leftskip 0pt\parindent 0pt\hsize=\largoo
\baselineskip 10pt
\ligne{\hfill
\arule 1  W W W W W R W 
\hfill}
\ligne{\hfill
\arule 1  W W W W R R W 
\hfill}
\ligne{\hfill
\arule 1  W W W W R B W 
\hfill}
\ligne{\hfill
\arule 1  W W B W R R W 
\hfill}
\ligne{\hfill
\arule 1  R W W W W R R 
\hfill}
\ligne{\hfill
\arule 1  B W W W W R B 
\hfill}
\ligne{\hfill
\arule 1  W W W W B R W 
\hfill}
\ligne{\hfill
\arule 1  R W W W W B R 
\hfill}
\ligne{\hfill
\arule 1  R W W W W W R 
\hfill}
\ligne{\hfill
\arule 1  W W W R W B W 
\hfill}
}

\setbox112=\vtop{\leftskip 0pt\parindent 0pt\hsize=\largoo
\baselineskip 10pt
\ligne{\hfill
\arule 1  R G B W W W R 
\hfill}
\ligne{\hfill
\arule 1  B G R W W W B 
\hfill}
\ligne{\hfill
\arule 1  B R R W W W B 
\hfill}
\ligne{\hfill
\arule 1  B G W W W R B 
\hfill}
\ligne{\hfill
\arule 1  W G W R W B W 
\hfill}
\ligne{\hfill
\arule 1  B R W B W W B 
\hfill}
\ligne{\hfill
\arule 1  B R B W W W B 
\hfill}
\ligne{\hfill
\arule 1  R B R W W W R 
\hfill}
\ligne{\hfill
\arule 1  R R B W W W R 
\hfill}
}

\setbox114=\vtop{\leftskip 0pt\parindent 0pt\hsize=\largoo
\baselineskip 10pt
\ligne{\hfill
\arule 1  W W B B G W G 
\hfill}
\ligne{\hfill
\arule 1  W G B W R R R 
\hfill}
\ligne{\hfill
\arule 1  G W B B W W W 
\hfill}
\ligne{\hfill
\arule 1  W G B B W W W 
\hfill}
\ligne{\hfill
\arule 1  W R B B W W W 
\hfill}
\ligne{\hfill
\arule 1  R W B W R R W 
\hfill}
\ligne{\hfill
\arule 1  B R W W W B B 
\hfill}
\ligne{\hfill
\arule 1  R G W W W W R 
\hfill}
}

\setbox116=\vtop{\leftskip 0pt\parindent 0pt\hsize=\largoo
\baselineskip 10pt
\ligne{\hfill
\arule 1  W W B B R W R 
\hfill}
\ligne{\hfill
\arule 1  R W B B W W W 
\hfill}
\ligne{\hfill
\arule 1  W R B W R R W 
\hfill}
\ligne{\hfill
\arule 1  W R W R W B G 
\hfill}
\ligne{\hfill
\arule 1  W W B B W G W 
\hfill}
\ligne{\hfill
\arule 1  G W W R W B W 
\hfill}
\ligne{\hfill
\arule 1  B G W B W W B 
\hfill}
\ligne{\hfill
\arule 1  B B R B W W B 
\hfill}
}

\vtop{
\vspace{-5pt}
\begin{tab}\label{rulcrossfix}
\leurre
The rules for the crossings and for the fixed switch. 
The first two columns are conservative rules. The last two columns contains the motion rules.
\end{tab}
\vspace{-10pt}
\ligne{\hfill\box110\hfill\box112\hfill\box114\hfill\box116\hfill}
}

\vskip 5pt
There is slightly less 
rules strictly required by the crossings than those used for the tracks themselves,
taking into account the above mentioned rules for the red locomotive. The rules
controlling the difference between a green locomotive and a red one are rules~72 and~76
for the cell~1(1) and rules~82 and~84 for the cell~1(5). After these cells, the cell~3(1)
for the red locomotive, and the cell~4(5) for the new green one, motion rules apply:
rule~39 for the cell~3(5) and rule~36 for the cell~4(5).

\subsubsection{Rules for the flip-flop}

   Table~\ref{rulflflp} gives the rules for the flip-flop, of course in its two versions.
The conservative rules are mainly required by the new state~\YY{} which was introduced
for the simulation of this switch. They are listed in the first column.

   Note that rule 105, which does not change the current state is typically a cell which 
contributes to the role of the switch. This rule applies to the cell~1(4) when the cell
sees the locomotive in the central cell. As the cell has a red neighbour a single white
neighbour, which is not the case of the cell~1(2) which, at the same time has two white
neighbours. This is why rule~104 prevents the locomotive to enter the first element of the
non selected track while rule~38 makes the locomotive enter the first element of the
selected track.

\setbox110=\vtop{\leftskip 0pt\parindent 0pt\hsize=\largoo
\baselineskip 10pt
\ligne{\hfill
\arule 1  G W Y G G W G 
\hfill}
\ligne{\hfill
\arule 1  Y G G B B B Y 
\hfill}
\ligne{\hfill
\arule 1  G G B W W W G 
\hfill}
\ligne{\hfill
\arule 1  G G W W W R G 
\hfill}
\ligne{\hfill
\arule 1  B G Y W W W B 
\hfill}
\ligne{\hfill
\arule 1  W W G R R R W 
\hfill}
\ligne{\hfill
\arule 1  G W R W W W G 
\hfill}
\ligne{\hfill
\arule 1  W W B W G R W 
\hfill}
\ligne{\hfill
\arule 1  G W R G G Y G 
\hfill}
\ligne{\hfill
\arule 1  R G W R R R R 
\hfill}
\ligne{\hfill
\arule 1  G G R W W W G 
\hfill}
\ligne{\hfill
\arule 1  G G W W W B G 
\hfill}
\ligne{\hfill
\arule 1  R R G W W W R 
\hfill}
\ligne{\hfill
\arule 1  R G R W W W R 
\hfill}
}

\setbox112=\vtop{\leftskip 0pt\parindent 0pt\hsize=\largoo
\baselineskip 10pt
\ligne{\hfill
\arule 1  W G W G W G G 
\hfill}
\ligne{\hfill
\arule 1  B W G B W B B 
\hfill}
\ligne{\hfill
\arule 1  G G W W W G W 
\hfill}
\ligne{\hfill
\arule 1  G G Y G G W G 
\hfill}
\ligne{\hfill
\arule 1  W G B W G R W 
\hfill}
\ligne{\hfill
\arule 1  G G R G G Y G 
\hfill}
\ligne{\hfill
\arule 1  W G G W W G W 
\hfill}
\ligne{\hfill
\arule 1  W G G R R R G 
\hfill}
\ligne{\hfill
\arule 1  G W Y G G G W 
\hfill}
\ligne{\hfill
\arule 1  W W G G G B W 
\hfill}
\ligne{\hfill
\arule 1  G W G R R R G 
\hfill}
\ligne{\hfill
\arule 1  R G G W W W R 
\hfill}
\ligne{\hfill
\arule 1  W W Y G G G W 
\hfill}
\ligne{\hfill
\arule 1  Y W G B B B R 
\hfill}
\ligne{\hfill
\arule 1  G W B W W W G 
\hfill}
}

\setbox114=\vtop{\leftskip 0pt\parindent 0pt\hsize=\largoo
\baselineskip 10pt
\ligne{\hfill
\arule 1  G W W R R R G 
\hfill}
\ligne{\hfill
\arule 1  W W R G G G B 
\hfill}
\ligne{\hfill
\arule 1  R W G B B B Y 
\hfill}
\ligne{\hfill
\arule 1  G W R G G R W 
\hfill}
\ligne{\hfill
\arule 1  B W Y G G G G 
\hfill}
\ligne{\hfill
\arule 1  Y B W B B B Y 
\hfill}
\ligne{\hfill
\arule 1  G B W W W R G 
\hfill}
\ligne{\hfill
\arule 1  G W B R R R R 
\hfill}
\ligne{\hfill
\arule 1  W W R G G Y G 
\hfill}
\ligne{\hfill
\arule 1  R W W R R R W 
\hfill}
\ligne{\hfill
\arule 1  G W Y G G R G 
\hfill}
\ligne{\hfill
\arule 1  W W R G W B W 
\hfill}
\ligne{\hfill
\arule 1  R W G R R R R 
\hfill}
\ligne{\hfill
\arule 1  G W W G G Y G 
\hfill}
}

\setbox116=\vtop{\leftskip 0pt\parindent 0pt\hsize=\largoo
\baselineskip 10pt
\ligne{\hfill
\arule 1  W G W R R R W 
\hfill}
\ligne{\hfill
\arule 1  G G Y G G R G 
\hfill}
\ligne{\hfill
\arule 1  W G R G W B W 
\hfill}
\ligne{\hfill
\arule 1  G G W G G Y G 
\hfill}
\ligne{\hfill
\arule 1  B G W B W B B 
\hfill}
\ligne{\hfill
\arule 1  W W B G G G W 
\hfill}
\ligne{\hfill
\arule 1  G W G G G Y W 
\hfill}
\ligne{\hfill
\arule 1  G G W R R R G 
\hfill}
\ligne{\hfill
\arule 1  Y G W B B B R 
\hfill}
\ligne{\hfill
\arule 1  W W G G G Y W 
\hfill}
\ligne{\hfill
\arule 1  R G W B B B Y 
\hfill}
\ligne{\hfill
\arule 1  W W G G G R B 
\hfill}
\ligne{\hfill
\arule 1  W W Y G G R G 
\hfill}
\ligne{\hfill
\arule 1  B W G G G Y G 
\hfill}
\ligne{\hfill
\arule 1  G B W R R R R 
\hfill}
}

\vtop{
\vspace{-5pt}
\begin{tab}\label{rulflflp}
\leurre
The rules for the flip-flop. 
%The first two columns are conservative rules. The last two columns contains the motion rules.
\end{tab}
\vspace{-10pt}
\ligne{\hfill\box110\hfill\box112\hfill\box114\hfill\box116\hfill}
\vskip 10pt
}

   When the flip-flop selects the track starting from~1(4), rule~35 applies to the cell. Note that
it is a rule already used for a simple motion of the locomotive on the tracks. Now, at
the same time, rule~132 applies to~1(2) which prevents the locomotive to enter the non-selected
track. Many rules are used to manage the change of the selection. Their action is symbolized by
Table~\ref{basc_exec} which shows in both positions of the switch the action of the rules in 
the cells concerned by the motion of the locomotive and the action of the switch.
As an example, the rules acting on the cell~2(1) when the selected track is the left-hand side one
are, successively: 88, 114, 118 and~121, the flash being controlled by rules~114 and~118. In the 
motion when the selected tracl is the right-hand side one, rules~88 and~121 again apply but the
flash is now controlled by rules~138 and~140.

\setbox110=\vtop{\leftskip 0pt\parindent 0pt\hsize=150pt
\obeylines
\obeyspaces\global\let =\ \footnotesize\tt
1(2) 2(2) 1(1) 2(1) 1(5) 2(5) 1(4)

  W    W    G    Y    G    R    W
  W    W    G    Y    G    R    W
  W    W    G    Y    G    R    W
  G    W    G    Y    G    R    W
  W    G    G    Y    G    R    W
  W    G    W    Y    G    R    W
  W    G    W    R    G    R    W
  W    G    B    Y    W    R    W
  W    R    G    Y    G    W    W
}

\setbox112=\vtop{\leftskip 0pt\parindent 0pt\hsize=150pt
\obeylines
\obeyspaces\global\let =\ \footnotesize\tt
1(2) 2(2) 1(1) 2(1) 1(5) 2(5) 1(4)

  W    R    G    Y    G    W    W
  W    R    G    Y    G    W    W
  W    R    G    Y    G    W    W
  W    R    G    Y    G    W    G
  W    R    G    Y    G    G    W
  W    R    G    Y    W    G    W
  W    R    G    R    W    G    W
  W    R    W    Y    B    G    W
  W    W    G    Y    G    R    W
}

\vtop{
\begin{tab}\label{basc_exec}
\leurre
Execution trace of the crossing of the flip-flop by the locomotive when the 
selected track is: to left, the left-hand side track; to right, the right-hand side track.
\end{tab}
\vspace{-5pt}
\ligne{\hfill\box110\hfill\box112\hfill}
}

\subsubsection{Rules for the memory switch}

   Here, the rules will be divided into two tables: Table~\ref{rulmemact} and 
Table~\ref{rulmempass} for the active and the passive part respectively.

   We know that the action of the locomotive crossing actively the active part of the memory
switch makes no change in the configuration of the switch. This is why the in first two columns
of Table~\ref{rulmemact} all rules keep the current state. In the last two columns, we can
see the rules involved by the flash of the switch triggered from the passive part of the
memory switch.

\vskip 5pt

\setbox110=\vtop{\leftskip 0pt\parindent 0pt\hsize=\largoo
\baselineskip 10pt
\ligne{\hfill
\arule 1  G W B G G W G 
\hfill}
\ligne{\hfill
\arule 1  G W R G G B G 
\hfill}
\ligne{\hfill
\arule 1  W W G G W B W 
\hfill}
\ligne{\hfill
\arule 1  W W W G G B W 
\hfill}
\ligne{\hfill
\arule 1  B G G B W B B 
\hfill}
\ligne{\hfill
\arule 1  B B G W W G B 
\hfill}
\ligne{\hfill
\arule 1  W B G W W G W 
\hfill}
\ligne{\hfill
\arule 1  W W G R B R W 
\hfill}
\ligne{\hfill
\arule 1  R G W R B R R 
\hfill}
}

\setbox112=\vtop{\leftskip 0pt\parindent 0pt\hsize=\largoo
\baselineskip 10pt
\ligne{\hfill
\arule 1  G G B G G W G 
\hfill}
\ligne{\hfill
\arule 1  G G R G G B G 
\hfill}
\ligne{\hfill
\arule 1  W G G R B R W 
\hfill}
\ligne{\hfill
\arule 1  G W B G G R G 
\hfill}
\ligne{\hfill
\arule 1  R W G R B R R 
\hfill}
\ligne{\hfill
\arule 1  G W W G G B G 
\hfill}
\ligne{\hfill
\arule 1  W G W R B R W 
\hfill}
\ligne{\hfill
\arule 1  G G B G G R G 
\hfill}
\ligne{\hfill
\arule 1  G G W G G B G 
\hfill}
}

\setbox114=\vtop{\leftskip 0pt\parindent 0pt\hsize=\largoo
\baselineskip 10pt
\ligne{\hfill
\arule 1  B G G B G B R 
\hfill}
\ligne{\hfill
\arule 1  G B G W W G W 
\hfill}
\ligne{\hfill
\arule 1  G W R G G W R 
\hfill}
\ligne{\hfill
\arule 1  R G G B W B B 
\hfill}
\ligne{\hfill
\arule 1  B R G W W G B 
\hfill}
\ligne{\hfill
\arule 1  W R G W W G W 
\hfill}
\ligne{\hfill
\arule 1  R W B G G W G 
\hfill}
\ligne{\hfill
\arule 1  B R W B W B B 
\hfill}
\ligne{\hfill
\arule 1  G R W W W R G 
\hfill}
}

\setbox116=\vtop{\leftskip 0pt\parindent 0pt\hsize=\largoo
\baselineskip 10pt
\ligne{\hfill
\arule 1  W W R R B R R 
\hfill}
\ligne{\hfill
\arule 1  W W R G G B G 
\hfill}
\ligne{\hfill
\arule 1  R W W R B R W 
\hfill}
\ligne{\hfill
\arule 1  G W W G G R R 
\hfill}
\ligne{\hfill
\arule 1  W W B G G R G 
\hfill}
\ligne{\hfill
\arule 1  B W R B W B B 
\hfill}
\ligne{\hfill
\arule 1  R W W G G B G 
\hfill}
\ligne{\hfill
\arule 1  W R W R B R R 
\hfill}
\ligne{\hfill
\arule 1  G R W W W B G 
\hfill}
}

\vtop{
\vspace{-15pt}
\begin{tab}\label{rulmemact}
\leurre
The rules for the active memory switch. First to columns to left: no change of the current state.
Last two columns: managing the flash of the switch.
\end{tab}
\vspace{-5pt}
\ligne{\hfill\box110\hfill\box112\hfill\box114\hfill\box116\hfill}
}
\vskip 5pt
\setbox110=\vtop{\leftskip 0pt\parindent 0pt\hsize=150pt
\obeylines
\obeyspaces\global\let =\ \footnotesize\tt
1(2) 2(2) 1(1) 2(1) 1(5) 2(5) 1(4)

  W    W    G    R    G    R    W
  W    W    R    B    W    R    W
  W    R    G    B    G    W    W
}
\setbox112=\vtop{\leftskip 0pt\parindent 0pt\hsize=150pt
\obeylines
\obeyspaces\global\let =\ \footnotesize\tt
1(2) 2(2) 1(1) 2(1) 1(5) 2(5) 1(4)

  W    R    G    R    G    W    W
  W    R    W    B    R    W    W
  W    W    G    B    G    R    W
}

\vtop{
\begin{tab}\label{flashmemact_exec}
\leurre
Execution trace of the flash in the active memory switch in both possible situations.
\end{tab}
\vspace{-5pt}
\ligne{\hfill\box110\hfill\box112\hfill}
}
\vskip 10pt
   Table~\ref{flashmemact_exec} produces the trace of execution when the flash issued from
the passive memory switch reaches the active memory one. The two kinds of actions are indicated.
The rules acting on the cell~2(1) are successively: 163, 166 and 170 when the selected track was
the left-hand side one. Rules 163 and 166 manage the flash and rule~170 witnesses the transfer
of the signal to the left-hand side. When the selected track is the rght-hand side one, rules~163
and~166 again manage the flash and this time, rule~177 witnesses that the signal goes to the 
right-hand side.

\vskip 10pt
We arrive to the passive memory switch which also requires a lot of rules, almost as many of them
as in the flip-flop, see Table~\ref{rulmempass}. The reason is the importance of the 
detection of the passage along the 
non-selected track. This triggers the change in the cell~0, which entails a change in both sides
of the cell. In the memory switch we apply the principle of changing the states when it is needed
to do so: a straightforward application of the definition of the memory switch could consist in
re-defining the selection even if the new selection is the same as the previous one. Here,
the change is triggered only if the locomotive passed through the non-selected track. This allows
us to implement a shortest process requiring less states.

\vskip 5pt
\setbox110=\vtop{\leftskip 0pt\parindent 0pt\hsize=\largoo
\baselineskip 10pt
\ligne{\hfill
\arule 1  B B W W Y Y B 
\hfill}
\ligne{\hfill
\arule 1  B B Y Y B W B 
\hfill}
\ligne{\hfill
\arule 1  Y B Y B W W Y 
\hfill}
\ligne{\hfill
\arule 1  Y B W W W W Y 
\hfill}
\ligne{\hfill
\arule 1  B Y W W W W B 
\hfill}
\ligne{\hfill
\arule 1  W Y W W W W W 
\hfill}
\ligne{\hfill
\arule 1  W Y Y W W W W 
\hfill}
\ligne{\hfill
\arule 1  W B Y W W W W 
\hfill}
\ligne{\hfill
\arule 1  Y B W B Y Y Y 
\hfill}
\ligne{\hfill
\arule 1  W Y W B W W W 
\hfill}
\ligne{\hfill
\arule 1  B Y W B W W B 
\hfill}
\ligne{\hfill
\arule 1  Y Y W W W W Y 
\hfill}
\ligne{\hfill
\arule 1  W Y B W W W W 
\hfill}
\ligne{\hfill
\arule 1  Y B Y R W Y Y 
\hfill}
\ligne{\hfill
\arule 1  Y Y Y W W W Y 
\hfill}
}

\setbox112=\vtop{\leftskip 0pt\parindent 0pt\hsize=\largoo
\baselineskip 10pt
\ligne{\hfill
\arule 1  R Y W W W W R 
\hfill}
\ligne{\hfill
\arule 1  W Y W B W B W 
\hfill}
\ligne{\hfill
\arule 1  W R Y W W W W 
\hfill}
\vskip 10pt
\ligne{\hfill
\arule 1  B B Y Y B G B 
\hfill}
\ligne{\hfill
\arule 1  B B G W Y Y B 
\hfill}
\ligne{\hfill
\arule 1  B G G W W W B 
\hfill}
\ligne{\hfill
\arule 1  W Y W B W G G 
\hfill}
\ligne{\hfill
\arule 1  B Y G B W W B 
\hfill}
\ligne{\hfill
\arule 1  Y B G B Y Y Y 
\hfill}
\ligne{\hfill
\arule 1  G Y W B W W W 
\hfill}
\ligne{\hfill
\arule 1  B B W G Y Y R 
\hfill}
\ligne{\hfill
\arule 1  W Y G B W W W 
\hfill}
\ligne{\hfill
\arule 1  R B W W Y Y B 
\hfill}
}

\setbox114=\vtop{\leftskip 0pt\parindent 0pt\hsize=\largoo
\baselineskip 10pt
\ligne{\hfill
\arule 1  B R Y Y B W Y 
\hfill}
\ligne{\hfill
\arule 1  W R W W G G W 
\hfill}
\ligne{\hfill
\arule 1  W R G B W W W 
\hfill}
\ligne{\hfill
\arule 1  Y R W B Y Y B 
\hfill}
\ligne{\hfill
\arule 1  Y R Y R W Y G 
\hfill}
\ligne{\hfill
\arule 1  B Y W W B G B 
\hfill}
\ligne{\hfill
\arule 1  Y Y G B W W Y 
\hfill}
\ligne{\hfill
\arule 1  B Y W W B W B 
\hfill}
\ligne{\hfill
\arule 1  W W Y W W B W 
\hfill}
\ligne{\hfill
\arule 1  G B Y R W Y Y 
\hfill}
\ligne{\hfill
\arule 1  Y G B W W W Y 
\hfill}
\ligne{\hfill
\arule 1  B Y W W B Y B 
\hfill}
\ligne{\hfill
\arule 1  Y Y Y B W W Y 
\hfill}
\ligne{\hfill
\arule 1  B Y G W W W B 
\hfill}
}

\setbox116=\vtop{\leftskip 0pt\parindent 0pt\hsize=\largoo
\baselineskip 10pt
\ligne{\hfill
\arule 1  Y B Y R G Y Y 
\hfill}
\ligne{\hfill
\arule 1  Y Y B W W W Y 
\hfill}
\ligne{\hfill
\arule 1  G Y W B W B W 
\hfill}
\ligne{\hfill
\arule 1  W G R W W W W 
\hfill}
\ligne{\hfill
\arule 1  B Y W G B Y B 
\hfill}
\ligne{\hfill
\arule 1  W W Y G W B G 
\hfill}
\ligne{\hfill
\arule 1  G W Y W W B W 
\hfill}
\ligne{\hfill
\arule 1  B Y G W B Y R 
\hfill}
\ligne{\hfill
\arule 1  W G Y W W B W 
\hfill}
\ligne{\hfill
\arule 1  R Y W W B Y B 
\hfill}
\ligne{\hfill
\arule 1  Y R Y Y B W B 
\hfill}
\ligne{\hfill
\arule 1  B R W B Y Y Y 
\hfill}
\ligne{\hfill
\arule 1  Y B G B W W Y 
\hfill}
\ligne{\hfill
\arule 1  Y G Y W W W Y 
\hfill}
}

\vtop{
\vspace{-15pt}
\begin{tab}\label{rulmempass}
\leurre
The rules for the passive memory switch. First column and upper part of the second one:
no change of the current state. Motion rules appear already in the lower part of the second
column. 
\end{tab}
\vspace{-5pt}
\ligne{\hfill\box110\hfill\box112\hfill\box114\hfill\box116\hfill}
}
\vskip 5pt

\setbox110=\vtop{\leftskip 0pt\parindent 0pt\hsize=150pt
\obeylines
\obeyspaces\global\let =\ \footnotesize\tt
1(2) 1(1) 0(0) 1(4) 1(3)

  W    B    B    Y    W
  W    B    B    Y    G
  W    B    R    Y    W
  W    Y    B    B    W
  W    Y    B    B    W
  W    Y    B    B    W
}
\setbox112=\vtop{\leftskip 0pt\parindent 0pt\hsize=150pt
\obeylines
\obeyspaces\global\let =\ \footnotesize\tt
1(2) 1(1) 0(0) 1(4) 1(3)

  W    Y    B    B    W
  G    Y    B    B    W
  W    Y    R    B    W
  W    B    B    Y    W
  W    B    B    Y    W
  W    B    B    Y    W
}

\vtop{
\begin{tab}\label{mempass_exec}
\leurre
Execution trace of the flash in the active memory switch in both possible situations.
\end{tab}
\vspace{-5pt}
\ligne{\hfill\box110\hfill\box112\hfill}
}

\vskip 10pt
Table~\ref{mempass_exec} allows us to see the process of changing the selected track when
the cell~0 can see the locomotive on the non-selected track. The rules which apply to this
cell are rules~181, 206, 208, 214 and 220{} when the selected track is the left-hand side one.
The flash is managed by rules~206 and~208, and rule~214 witnesses the change of selection. 
Rule~220 is the conservative rule for the other position. Namely, when the selected track is
the right-hand side one, the rules are 220, 230, 232, 203 and~181. New rules manage the flash
as the cell~0(0) can directly see from where the locomotive comes. Rule~215 again witnesses
the change and the conservative rule is again rule~181.
  
\vskip 5pt
   With these rules and with this study illustrated bu the figures, we completed the proof
of the following result:

\begin{thm}\label{univ5}
There is a rotation invariant cellular automaton on the pentagrid with $5$~states which is
planar and weakly universal.
\end{thm}
%\acknowledgements{...}
%\bibliographystyle{eptcs}
%\bibliography{submission_mm}

\end{document}